\documentclass[12pt]{article}
\usepackage{amsmath}
\usepackage{amssymb}

\usepackage[dvips]{color}
\renewcommand{\overline}[1]{\bar{#1}}
\makeatletter\@addtoreset{equation}{section}
\makeatother
\begin{document}
\begin{titlepage}
\begin{flushright}
TIT/HEP-572\\
HIP-2007-29/TH\\
May, 2007\\
\end{flushright}
\vspace{0.5cm}
\begin{center}
{\Large \bf 
Deformation of Super Yang-Mills Theories\\
in R-R 3-form Background 
}
\lineskip .75em
\vskip1.0cm
{\large Katsushi Ito${}^{1}$, Hiroaki Nakajima${}^{2}$ and Shin 
Sasaki${}^{3}$ }
\vskip 2.5em
${}^{1}$ {\normalsize\it Department of Physics\\
Tokyo Institute of Technology\\
Tokyo, 152-8551, Japan}  \vskip 1.5em
${}^{2}$ {\normalsize\it Department of Physics and Institute of Basic Science\\
Sungkyunkwan University\\
Suwon, 440-746, Korea
} \vskip 1.5em
${}^{3}$ {\normalsize\it High Energy Physics Division, Department of Physical 
Sciences\\
Helsinki Institute of Physics\\
University of Helsinki\\
P.O.Box 64, FIN-00014, Finland 
} 
\vskip 4.5em
\end{center}
\begin{abstract}
We study deformation of ${\cal N}=2$ and ${\cal N}=4$ 
super Yang-Mills theories,  which are obtained as the 
 low-energy effective theories on the (fractional) D3-branes 
in the presence of constant Ramond-Ramond 3-form background. 
We calculate the Lagrangian at the second order in the deformation
 parameter from open string disk amplitudes.
In ${\cal N}=4$ case we find that all supersymmetries are broken for generic
deformation parameter but part of supersymmetries are unbroken for 
special case.
We also find that classical vacua admit fuzzy sphere configuration.
In ${\cal N}=2$ case we determine the deformed supersymmetries. 
We rewrite the deformed Lagrangians 
in terms of ${\cal N}=1$ superspace,
where the deformation is interpreted as that of coupling constants.
\end{abstract}
\end{titlepage}

\baselineskip=0.7cm

\tableofcontents

\section{Introduction}
Low-energy effective field theories on D-branes in closed string
backgrounds have attracted much attentions.
The effects of the Ramond-Ramond (R-R) backgrounds are particularly 
interesting for studying (non-)perturbative properties of supersymmetric
gauge theories and superstrings.
For example, the constant graviphoton background,
which comes from the self-dual R-R 5-form field strenth wrapping three
cycle in a Calabi-Yau manifold,
produces 
stringy corrections to the F-terms in effective
theories \cite{AnGaNaTa, BeCeOoVa}. 
Such corrections play an important role 
in studying non-perturbative  properties of supersymmetric
gauge theories \cite{Ne, DiVa}.
Closed string background is also interesting from the geometrical
point of view because it deforms the world-volume geometry of D-branes.
A well-known example is the constant NS-NS $B$-field, which leads to 
the noncommutative space-time realized by the Moyal product \cite{ChuHo, SeWi}.
The R-R background also deforms the world-volume geometry.
In fact, the constant self-dual R-R 5-form background on the fractional 
D3-branes 
introduces non(anti)commutativity of 
(Euclidean) 
superspace \cite{OoVa, BeSe, DeGrNi}.
The deformed supersymmetric gauge theories on non(anti)commutative
superspace are studied extensively 
\cite{Se, ArItOh, FeIvLeSoZu, CaIvLeQu, SaWo}.

Since superstring theory contains R-R fields, 
it would be interesting to study deformed supersymmetric gauge theories 
in R-R backgrounds with various ranks 
and their (non-)perturbative properties.
Recently Bill\'o {\it et al} \cite{BiFrFuLe} 
studied 
the effective action on the fractional D3-D$(-1)$ system
in the R-R 3-form background ${\cal F}$ with fixed
$(2\pi\alpha')^{1/2}{\cal F}$ in the zero slope limit.
They showed that the deformed action agrees with the 
instanton effective action of ${\cal N}=2$ supersymmetric
gauge theory  in the $\Omega$-background \cite{Ne} 
at the lowest order of the 
deformation parameter and gauge coupling constant.
The $\Omega$-background utilize the integral over the instanton moduli space 
\cite{Ne, LoMaNe}. 
This type of deformation is not obtained from the non(anti)commutative
deformation of superspace.
It is an interesting problem to study geometrical meaning of this deformation.

In order to examine
the effects of R-R background, 
the most  direct approach is to 
calculate the 
low-energy effective action on the D-branes from superstring theory.
One can compute the action of non(anti)commutative gauge theories 
directly from the effective action on the (fractional) D3-branes
\cite{BiFrPeLe, BiFrFuLe, ItSa, ItKoSa},
where interaction terms are obtained from the open string 
disk amplitudes with insertion of graviphoton vertex operators.
For example, the deformed action of ${\cal N}=1$ supersymmetric 
gauge theories
was derived from the fractional D3-branes in type IIB superstring
theories compactified on ${\bf C}^3/{\bf Z}_2\times {\bf Z}_2$
\cite{BiFrPeLe}.
The effective theory
is ${\cal N}=1$ super Yang-Mills theory on non(anti)commutative 
${\cal N}=1$ superspace \cite{Se}.

In \cite{ItSa, ItKoSa} we discussed the deformation of ${\cal N}=2$ and
${\cal N}=4$ super Yang-Mills
theories in the R-R background field strength 
of the form $\mathcal{F}^{\alpha \beta AB}$, where $\alpha$ and $\beta$ label
the spinor indices of (Euclidean) space-time and $A$ and $B$ are internal
spinor indices.
We classify the field strength into four
types $\mathcal{F}^{(\alpha \beta) (AB)}$, $\mathcal{F}^{[\alpha \beta]
(AB)}$, $\mathcal{F}^{(\alpha \beta) [AB]}$ and $\mathcal{F}^{[\alpha
\beta] [AB]}$. 
Here $(ab)$ ($[ab]$) denotes the (anti)symmetrization of $ab$.
We call these deformations  as
(S,S), (A,S), (S,A) and (A,A)-type, respectively, where 
the (S,S)-type deformation  with fixed 
$(2\pi\alpha')^{3/2}{\cal F}$ 
corresponds to the case studied in \cite{BiFrPeLe}.
In \cite{ItSa}, we studied the first order correction to ${\cal N}=2$ 
super Yang-Mills action from the (S,S)-type background with fixed 
$(2\pi\alpha')^{3/2}{\cal F}$.
We showed that deformed theory agrees with ${\cal N}=2$ super
Yang-Mills theory on non(anti)commutative ${\cal N}=2$ harmonic
superspace \cite{ArItOh, FeIvLeSoZu, CaIvLeQu}.
In \cite{ItKoSa}, we studied the first order correction to ${\cal N}=4$ 
super Yang-Mills theory in (S,S)-type background with fixed 
$(2\pi\alpha')^{3/2}{\cal F}$.
By restricting the deformation parameter to the special case, 
the deformed Lagrangian is reduced to the one in non(anti)commutative 
${\cal N}=1$ superspace.
Therefore it is natural to think that the 
(S,S)-type deformation with fixed $(2\pi\alpha')^{3/2}{\cal F}$
corresponds to the non(anti)commutative deformation of ${\cal N}(\leq
4)$ extended
superspace at full order in deformation parameter.
On the other hand, the index structure of the (A,A) type background 
suggests that it corresponds to the singlet deformation of extended 
superspace \cite{FeIvLeSoZu, ArIt}, although we need to take into account 
the backreaction to the closed string backgrounds \cite{ItSa}.
The (S,A) and (A,S) type deformations with fixed $(2\pi\alpha')^{3/2}{\cal F}$ 
would also provide nontrivial deformation of supersymmetric gauge theories, 
which cannot be realized as non(anti)commutative superspace. 
However, it is difficult to compute
the deformed actions due to its complicated structure.

As shown in \cite{BiFrFuLe}, 
the (S,A)-type background with fixed $(2\pi\alpha')^{1/2}{\cal F}$
provides nontrivial deformation of ${\cal N}=2$ super Yang-Mills theory,
which is useful for studying instanton calculus.
Hence it would be an interesting problem to work out the 
deformations by the constant R-R backgrounds with fixed 
$(2\pi \alpha')^{1/2}{\cal F}$ and their non-perturbative properties.
The purpose of this paper is to study the deformation of super
Yang-Mills theories with ${\cal N}=2$ and $4$ supersymmetries
corresponding to the (S,A) and (A,S)-types background with
fixed $(2\pi\alpha')^{1/2}{\cal F}$.

We will calculate disk amplitudes with one R-R vertex operator
and derive the effective action on the (fractional) D-branes.
For ${\cal N}=4$ case, 
we will show that the bosonic action agrees with the Chern-Simons action
with the (dual) R-R potentials \cite{My}.
The deformed scalar potential has nontrivial minima. 
Actually, for both (S,A) and (A,S)-type deformations of 
$ \mathcal{N} = 4$ super Yang-Mills theory,
we find a fuzzy sphere configuration \cite{Im,ItKoSa} for adjoint scalars.
In general number of unbroken 
supersymmetries are restricted on the D-branes in 
the presence of R-R backgrounds.
We will examine invariance of the deformed Lagrangian under remaining
supersymmetries. 
The deformation of ${\cal N}=2$ super Yang-Mills theory is 
obtained from ${\cal N}=4$ theory 
by the reduction due to the ${\bf Z}_2$ orbifold of ${\bf C}^2$. 
For both ${\cal N} = 2$ and ${\cal N} = 4$ cases, we are able to explore geometrical interpretation of 
this deformation in terms of superspace formalism.
We will show that (S,A) and (A,S)-type deformations with fixed
$(2\pi \alpha')^{1/2}{\cal F}$ are realized by
introducing superspace dependent coupling constants.
This is in contrast with the case with the (S,S)-type deformation with 
fixed $(2\pi \alpha')^{3/2}{\cal F}$, where its deformation is 
realized by the star product for supercoordinates.

This paper is organized as follows: In section \ref{N4deformation}, 
we calculate the (S,A) and (A,S)-type background corrections to 
$\mathcal{N} = 4$ 
super Yang-Mills theory from the open string disk amplitudes 
with one closed string R-R vertex operator.
Unbroken supersymmetries are classified in terms of
the rank of deformation parameter in some cases.
The fuzzy sphere configurations of vacuum in the 
deformed theories are investigated. 
In section \ref{NACS}, 
we confirm that 
the R-R correction terms in (S,A) and (A,S)-type deformed
${\cal N}=4$ theories are consistent with the Chern-Simons term of the 
D-brane effective action coupled to
the R-R potential.
In section \ref{N2deformation}, we study the (S,A) and (A,S)-type 
deformations of $\mathcal{N} = 2$ super Yang-Mills theory 
and its deformed supersymmetry.
In section \ref{superspace} 
we rewrite the deformed action in terms of ${\cal N}=1$ superspace
and show that
(A,S)-type deformation is regarded as 
the mass deformation of super Yang-Mills theory.
Section \ref{discussion} is devoted to conclusions and discussion.

\section{Deformed $\mathcal{N}=4$ Super Yang-Mills theory in R-R 3-form
 background \label{N4deformation}}

In this section we study the low-energy effective action 
on D3-branes in type IIB superstrings from the 
disk amplitudes with one R-R vertex operator of (S,A) or (A,S)-type. 
Here we use NSR formalism and introduce spin fields 
\cite{FrMaSh, KoLeLeSaWa} to represent space-time spinor. 
The low-energy effective field theory on $N$ D3-branes are described by
gauge fields $A_\mu$ ($\mu=1,2,3,4$), 
six real scalars $\varphi^a$ ($a=5,\ldots,10$) and 
Weyl fermions $\Lambda_{\alpha} {}^A$ and
$\overline{\Lambda}^{\dot{\alpha}} {}_{A}$ ($A=1,2,3,4$), 
which belong to the adjoint representation of gauge group $U(N)$ .
We denote $T^m$ as the basis of $U(N)$ generators normalized as
${\rm Tr}(T^m T^n)=k\delta^{mn}$ with constant factor $k$.

The vertex operators for these fields are \cite{BiFrPeFuLeLi}
\begin{eqnarray}
V_A^{(-1)} (y;p) &=& (2 \pi \alpha')^{\frac{1}{2}} \frac{A_{\mu} 
 (p)}{\sqrt{2}} \psi^{\mu} (y) e^{- \phi (y)} e^{i \sqrt{2 \pi \alpha'} p
 \cdot X (y)}, \nonumber \\
V_A^{(0)} (y;p) &=& 2 i (2 \pi \alpha')^{\frac{1}{2}} A_{\mu} (p) 
 \left( \partial X^{\mu} (y) + i (2 \pi \alpha')^{\frac{1}{2}} p \cdot 
  \psi \psi^{\mu} (y)  \right) e^{i \sqrt{2 \pi \alpha'} p \cdot X (y)}.
\end{eqnarray}
\begin{eqnarray}
V_{\varphi}^{(-1)} (y;p) &=& (2 \pi \alpha')^{\frac{1}{2}} \frac{\varphi_a 
 (p)}{\sqrt{2}} \psi^a (y) e^{- \phi (y)} e^{i \sqrt{2 \pi \alpha'} p
 \cdot X (y)}, \nonumber \\
V_{\varphi}^{(0)} (y;p) &=& 2 i (2 \pi \alpha')^{\frac{1}{2}} \varphi_a (p) 
 \left( \partial X^a (y) + i (2 \pi \alpha')^{\frac{1}{2}} p \cdot 
  \psi \psi^a (y)  \right) e^{i \sqrt{2 \pi \alpha'} p \cdot X (y)}.
\end{eqnarray}
\begin{eqnarray}
V_{\Lambda}^{(-1/2)} (y;p) &=& \frac{}{} (2 \pi \alpha')^{\frac{3}{4}} 
 \Lambda^{\alpha A} (p) S_{\alpha} (y) S_A (y) e^{- \frac{1}{2} \phi 
 (y)} e^{i \sqrt{2 \pi \alpha' } p \cdot X (y)}, \nonumber \\
V_{\overline{\Lambda}}^{(-1/2)} (y;p) &=& \frac{}{}( 2 \pi \alpha' 
 )^{\frac{3}{4}} \overline{\Lambda}_{\dot{\alpha} A} (p) S^{\dot{\alpha}} (y) 
 S^A (y) e^{-\frac{1}{2} \phi (y)} e^{i \sqrt{2 \pi \alpha' } p 
 \cdot X (y)}.
\end{eqnarray}
Here $(X^{M}(z),\psi^{M}(z))$ ($M=1,\ldots,10$) 
are free bosons and
fermions on the worldsheet, where $\mu$ labels the worldvolume
coordinates on D3-branes and $a$ 
coordinates transverse to the worldvolume of the D3-branes. 
$S_\alpha$ and $S_A$ denote the spin operators for space-time and
internal space parts.
$\phi$ is a free boson obtained from the bosonization of
the bosonic ghost $(\beta,\gamma)$.
For gauge fields and scalar fields we use two physically equivalent
 vertex operators with picture number $-1$ and $0$.
For fermions we use the vertex operator with picture number $-1/2$.

The disk amplitudes in the zero-slope limit $\alpha'\rightarrow 0$ 
reproduce the action of ${\cal N}=4$ super Yang-Mills theory.
It is convenient to introduce auxiliary field vertex operators
in order to reduce higher point amplitudes to
the lower ones \cite{DiIcSe, BiFrPeFuLeLi, ItSa, ItKoSa}.
These are given by
\begin{eqnarray}
& & V_{H_{AA}}^{(0)} (y;p) = {1\over2} (2 \pi \alpha') H_{\mu \nu} (p) 
 \psi^{\mu} \psi^{\nu} (y) e^{i \sqrt{2 \pi \alpha'} p \cdot X (y)},
\nonumber \\
& & V_{H_{A \varphi}}^{(0)} (y;p) = 2 (2 	\pi \alpha') H_{\mu a} (p) 
 \psi^{\mu} \psi^{a} (y) e^{i \sqrt{2 \pi \alpha'} p \cdot X (y)},
\nonumber \\
& & V_{H_{\varphi \varphi}}^{(0)} (y;p) = -{1\over\sqrt{2}}
 (2 \pi \alpha') H_{ab} (p) 
 \psi^{a} \psi^{b} (y) e^{i \sqrt{2 \pi \alpha'} p \cdot X (y)}.
\end{eqnarray}
Note that these vertex operators are not BRST invariant.
The total Lagrangian includes only the cubic interaction terms and 
becomes
\begin{eqnarray}
\mathcal{L}_{\mathcal{N} = 4}
 &=&  
-  \frac{1}{k g^2_{\mathrm{YM}}} 
 \mathrm{Tr} \left[ \frac{1}{2} (\partial_{\mu} A_{\nu} - \partial_{\nu} 
A_{\mu}) \partial^{\mu} A^{\nu} +  i \partial_{\mu} A_{\nu} [A^{\mu}, A^{\nu}]
+ \frac{1}{2} H_c H^c + \frac{1}{2} H_c \eta^c_{\mu \nu} [A^{\mu},
A^{\nu}] 
\right] 
\nonumber \\
& &  -  \frac{1}{k g^2_{\mathrm{YM}}} \mathrm{Tr} 
\left[ \frac{1}{2} H_{ab} H_{ab} + \frac{1}{\sqrt{2}} H_{ab} 
[\varphi_a, \varphi_b] \right] 
\nonumber \\
& &  -  \frac{1}{k g^2_{\mathrm{YM}}} \mathrm{Tr} \left[ \frac{1}{2} 
\partial_{\mu} 
\varphi_a \partial^{\mu} \varphi_a + i \partial_{\mu} \varphi_a
[A^{\mu}, 
\varphi_a] 
+ \frac{1}{2} H_{\mu a} {H}^{\mu a} +  H_{\mu a} [A^{\mu}, \varphi_a]  
\right] 
\nonumber \\
& &  - \frac{1}{k g^2_{\mathrm{YM}}} \mathrm{Tr} \left[ i \Lambda^A 
\sigma^{\mu} 
D_{\mu} \bar{\Lambda}_A - \frac{1}{2} \left( \Sigma^a \right)^{AB} 
\overline{\Lambda}_{\dot{\alpha}A}
[\varphi_a, \overline{\Lambda}^{\dot{\alpha}}_{\ B} ] - \frac{1}{2} \left( 
\overline{\Sigma}^a \right)_{AB} \Lambda^{\alpha A} [\varphi_a, 
\Lambda_{\alpha}^B ]  \right]. \nonumber \\
\end{eqnarray}
Here the four-dimensional Euclidean sigma matrices are
$\sigma_\mu=(i\tau^1,i\tau^2,i\tau^3,1)$ and
$\bar{\sigma}_\mu=(-i\tau^1,-i\tau^2,-i\tau^3,1)$,
where $\tau^i$ ($i=1,2,3$) are the Pauli matrices.
The six-dimensional sigma matrices are 
given by
\begin{equation}
\Sigma^a=
\left( \eta^{3}, - i\bar{\eta}^{3}, \eta^{2}, - i\bar{\eta}^{2},
\eta^{1}, i\bar{\eta}^{1} \right),
\quad \bar{\Sigma}^a=
(-\eta^3, -i\bar{\eta}^3, -\eta^2,-i\bar{\eta}^2, -\eta^1, i\bar{\eta}^1), 
\label{eq:sixgamma}
\end{equation}
where $a=5,\cdots,10$. 
$\eta^{i}_{\mu\nu}$ and $\bar{\eta}^{i}_{\mu\nu}$ are
{}'t Hooft symbols, which are defined by
$\sigma_{\mu\nu}={i\over2}\eta^{i}_{\mu\nu} \tau^{i}$ and
$\bar{\sigma}_{\mu\nu}={i\over2}\bar{\eta}^{i}_{\mu\nu} \tau^{i}$.
After integrating out the auxiliary fields, we get 
the quartic interaction terms including the gauge fields and scalars,
which is given by
\begin{eqnarray}
\mathcal{L}^{(0)}_{\mathcal{N}=4}
 &=&  
\frac{1}{k g^2_{\mathrm{YM}}} \mathrm{Tr} \left[ 
- \frac{1}{4} F^{\mu\nu} \left(F_{\mu \nu} + \tilde{F}_{\mu \nu} \right) 
-i \Lambda^{\alpha A} (\sigma^{\mu})_{\alpha \dot{\beta}} D_{\mu} 
\overline{\Lambda}^{\dot{\beta}}_{\ A}
    - \frac{1}{2} \left(D_{\mu} \varphi_a \right)^2 \right. \nonumber \\
& & \left. \quad \quad + \frac{1}{2} 
\left( \Sigma^a \right)^{AB} \overline{\Lambda}_{\dot{\alpha}A}
[\varphi_a, \overline{\Lambda}^{\dot{\alpha}}_{\ B} ] + \frac{1}{2} \left( 
\overline{\Sigma}^a \right)_{AB} \Lambda^{\alpha A} [\varphi_a, 
\Lambda_{\alpha}^B ] + \frac{1}{4} [\varphi_a, \varphi_b]^2   \right].
\label{undef}
\end{eqnarray}
We call ${\cal L}^{(0)}_{\mathcal{N} = 4}$ undeformed Lagrangian.

We then introduce a R-R closed string vertex operator 
\begin{eqnarray}
V_{\mathcal{F}}^{(-1/2,-1/2)} (z, \bar{z}) = (2 
 \pi \alpha') \mathcal{F}^{\alpha \beta AB} 
 \left[ S_{\alpha} (z)  S_A (z) e^{- \frac{1}{2} \phi (z)} 
  S_{\beta} (\bar{z})  S_B (\bar{z}) e^{ - \frac{1}{2} 
  \phi (\bar{z})} \right]
\label{eq:rrvertex1}
\end{eqnarray}
with constant $\mathcal{F}^{\alpha \beta AB}$ and
insert this vertex operator in a disk amplitude.
Here we have used 
the doubling trick for the spin fields in
(\ref{eq:rrvertex1}) and have replaced
the right-moving part in the R-R vertex operator by 
$S_{\beta} (\bar{z})  S_B (\bar{z}) e^{ - \frac{1}{2} 
  \phi (\bar{z})}$.
The disk amplitude is now given by
\begin{equation}
 \langle \! \langle V^{(q_1)}_{X_1}\cdots V^{(-{1\over2},-{1\over2})}_{\cal
  F}
\cdots \rangle \! \rangle
=C_{D_2}\int {\prod_{i=1}^{n}dy_i \prod_{j=1}^{n_{\cal F}}
dz_jd\bar{z}_j
\over dV_{CKG}}
\langle V^{(q_1)}_{X_1}(y_1)\cdots V^{(-{1\over2},-{1\over2})}_{\cal F}
(z_1,\bar{z_1})\cdots \rangle,
\label{eq:disk1}
\end{equation}
where 
$V^{(q_k)}_{X_k}$ is the open string vertex operator corresponding to 
a field $X_k$ with picture number $q_k$, 
$C_{D_2}={1\over 2\pi^2(\alpha')^2 }{1\over k g_{\mathrm{YM}}^2}
$ is a normalization factor and $dV_{CKG}$ is an 
$SL(2,{\bf R})$-invariant volume factor to fix 
positions of three coordinates in $y_i$, $z_j$ and
$\bar{z}_j$. 
The sum of picture numbers in a disk amplitude must be $-2$.

The constant R-R field strength $\mathcal{F}^{\alpha \beta AB} $ 
is decomposed into the types
$\mathcal{F}^{(\alpha \beta) (AB)}$, $\mathcal{F}^{[\alpha \beta]
(AB)}$, $\mathcal{F}^{(\alpha \beta) [AB]}$ and $\mathcal{F}^{[\alpha
\beta] [AB]}$, which are called (S,S), (A,S), (S,A) and (A,A)-type, 
respectively.
It is shown in \cite{ItKoSa} that the (S,S)-type background corresponds to
the R-R 5-form and the (A,S) and (S,A)-types to 3-forms and its dual 7-forms, 
the (A,A)-type to the 1-form and its dual 9-form.
In order to discuss the zero-slope limit, we need to specify the scaling 
condition for ${\cal F}$.
In the previous paper \cite{ItKoSa}, we have studied the (S,S)-type
deformation with the scaling condition $(2\pi\alpha)^{3/2}{\cal F}$
fixed, which would correspond to the deformation of underlying 
${\cal N}=4$ extended superspace.

In this paper we will consider the (S,A) and (A,S)-type deformations
with different scaling condition $(2\pi \alpha')^{1/2}{\cal F}$ fixed.
These types of deformations cannot be realized by introducing
non(anti)commutativity of superspace and give new types
of deformed theories.
The scaling condition ${\cal F}\sim (\alpha')^{-1/2}$ is particularly
interesting because it provides the (S,A)-type 
deformation of D$(-1)$-instanton
effective action similar to the $\Omega$-background in
${\cal N}=2$ super Yang-Mills theory \cite{Ne, BiFrFuLe}.
We will consider the effects of the 
R-R 3-form field strength of (S,A) and (A,S)-types
 to the low-energy effective Lagrangian 
in the ${\cal N}=4$ case.

\subsection{(S,A)-type deformation}
\subsubsection{Lagrangian}
Firstly we discuss the (S,A)-type deformation 
of ${\cal N}=4$ super Yang-Mills theory.
For the (S,A)-type background ${\cal F}^{(\alpha\beta)[AB]}$, 
we find that the disk amplitudes which are nonzero in the zero-slope
limit are given by
$\langle \! \langle V_A V_{\varphi} V_{\mathcal{F}} \rangle \! \rangle$, 
 $\langle \! \langle V_{H_{AA}} V_{\varphi} V_{\mathcal{F}} 
\rangle \! \rangle$ 
 and $ \langle \! \langle V_{\Lambda} 
 V_{\Lambda} V_{\mathcal{F}} \rangle \! \rangle$.
The explicit computations of these amplitudes are essentially the same
as in \cite{ItKoSa}. 
We do not repeat detailed calculations here.
The first two amplitudes become
\begin{eqnarray}
& & \langle \! \langle V^{(0)}_A (p_1) V_{\varphi}^{(-1)} 
(p_2) V^{(-1/2,-1/2)}_{\mathcal{F}} \rangle \! 
 \rangle \nonumber \\
& & =  - (-i) \frac{4 \pi}{k g^2_{\mathrm{YM}}} \mathrm{Tr}
\left[ (\sigma^{\mu \nu})_{\alpha \beta} i p_{1 \mu} A_{\nu} (p_1) 
(\overline{\Sigma}^a)_{AB} \varphi_a (p_2) 
\right] (2 \pi \alpha')^{\frac{1}{2}} \mathcal{F}^{(\alpha \beta)[AB]},
\end{eqnarray}
and
\begin{eqnarray}
& & \langle \! \langle V^{(0)}_{H_{AA}} (p_1) V_{\varphi}^{(-1)} 
(p_2) V^{(-1/2,-1/2)}_{\mathcal{F}} \rangle \! 
 \rangle \nonumber \\
& & = - (-i) \frac{1}{2i} \frac{1}{2} \frac{4 \pi}{k g^2_{\mathrm{YM}}}
 \mathrm{Tr} \left[ (\sigma^{\mu \nu})_{\alpha \beta} H_{\mu \nu} (p_1) 
(\overline{\Sigma}^a)_{AB} \varphi_a (p_2) 
\right] (2 \pi \alpha')^{\frac{1}{2}} \mathcal{F}^{(\alpha \beta)[AB]} 
.
\end{eqnarray}
The interaction terms corresponding to  these amplitudes are given by
\begin{eqnarray}
 - \frac{2 \pi i}{k g^2_{\mathrm{YM}}}
\mathrm{Tr} \left[ (\sigma^{\mu \nu})_{\alpha \beta} 
\left( \partial_{[ \mu } A_{\nu]} - \frac{i}{2} H_{\mu \nu}  \right)
(\overline{\Sigma}^a)_{AB} \varphi_a \right]
(2 \pi \alpha')^{\frac{1}{2}} \mathcal{F}^{(\alpha \beta)[AB]}.
\label{eq:asint1}
\end{eqnarray}
The third amplitude is 
\begin{eqnarray}
& & \langle \! \langle V^{(-1/2)}_{\Lambda} (p_1) V^{(-1/2)}_{\Lambda} (p_2) V^{(-1/2,-1/2)}_{\mathcal{F}}
\rangle \! \rangle \nonumber \\
& & = i \frac{4 \pi i}{k g^2_{\mathrm{YM}}}
\mathrm{Tr} \left[ \varepsilon_{ABCD} \Lambda_{\alpha}^{\ A} (p_1) 
\Lambda_{\beta}^{\ B} (p_2) \right] (2 \pi \alpha')^{\frac{1}{2}} 
\mathcal{F}^{(\alpha \beta)[CD]}.
\label{eq:third}
\end{eqnarray}
Introducing symmetric factor in (\ref{eq:third}) and 
adding the terms (\ref{eq:asint1}), we obtain the interaction term 
including auxiliary fields.
Integrating out 
the auxiliary fields, we find that the deformed Lagrangian is 
$\mathcal{L}^{(0)}_{\mathcal{N} = 4}
+\mathcal{L}^{(1)}_{\mathrm{(S,A)}}
+\mathcal{L}^{(2)}_{\mathrm{(S,A)}}+\cdots$, where
\begin{eqnarray}
\mathcal{L}^{(1)}_{\mathrm{(S,A)}}
 &=& \frac{1}{k g^2_{\mathrm{YM}}}
\mathrm{Tr} \left[i F_{\mu \nu} \varphi_a \right] C^{\mu \nu a} 
- \frac{1}{k g^2_{\mathrm{YM}}}
\mathrm{Tr} \left[ \varepsilon_{ABCD} \Lambda_{\alpha}^{\ A}
	     \Lambda_{\beta}^{\ B}  \right] C^{(\alpha \beta)[CD]},
\label{N4sa1}\\
\mathcal{L}^{(2)}_{\mathrm{(S,A)}}
&=& \frac{1}{2} \frac{1}{k g^2_{\mathrm{YM}}} \mathrm{Tr} \left[\varphi_a 
 \varphi_b  \right] C_{\mu \nu}^{\ \ a} C^{\mu \nu b}.
\label{N4sa2}
\end{eqnarray}
Here we have defined the deformation parameter by
\begin{eqnarray}
C^{\mu \nu a} &\equiv& - 2 \pi (2 \pi \alpha')^{\frac{1}{2}} (\sigma^{\mu 
 \nu})_{\alpha \beta} \left(\overline{\Sigma}^a \right)_{AB} 
 \mathcal{F}^{(\alpha \beta) [AB]}, \nonumber \\
C^{(\alpha \beta)[AB]} &\equiv& - 2 \pi (2 \pi \alpha')^{\frac{1}{2}}
\mathcal{F}^{(\alpha \beta) [AB]}.
\end{eqnarray}
The $O(C^2)$ term 
${\cal L}_{\mathrm{(S,A)}}^{(2)}$ arises from the integration 
over the auxiliary field. 
It is possible to construct higher order $O(C^n)$ terms
{}from the disk amplitudes.
It is not clear that these amplitudes are reducible or not.
For example, at order $C^2$, there is an amplitude
$\langle \! \langle V_{H_{\varphi \varphi}} V_{\mathcal{F}} 
V_{\mathcal{F}} \rangle \! \rangle$,
which might change the coefficients of the $\varphi^2 C^2$ term 
in ${\cal L}^{(2)}_{\mathrm{(S,A)}}$.
However, as we will see in section 4, 
the reduction from ${\cal N}=4$ to ${\cal N}=2$ theory shows that
${\cal L}^{(2)}_{\mathrm{(S,A)}}$ gives the $O(C^2)$ term of the ${\cal
N}=2$ theory, 
where the $O(C^2)$ term is exact. 
Moreover, as we see in the next subsection, the $O(C^2)$ deformed 
Lagrangian is invariant under $O(C)$ deformed supersymmetry for some 
$C$. 
This is rather different from non(anti)commutative ${\cal N}=2$
supersymmetric gauge theory, where deformed supersymmetry transformation 
contains higher order contributions of the deformation parameter \cite{ArItOh2}.
In order to cancel this deformation transformation, it is necessary to 
introduce infinite number of interaction terms.
But for the (S,A)-deformed Lagrangian we do not need to introduce such a 
higher order counter term.
These properties suggest that the deformed Lagrangian 
${\cal L}^{(0)}_{\mathcal{N} = 4}
+{\cal L}^{(1)}_{\mathrm{(S,A)}}+{\cal L}^{(2)}_{\mathrm{(S,A)}}$ is an exact 
Lagrangian, which would be difficult to prove in the NSR formalism.

\subsubsection{Deformed Supersymmetry}
We examine supersymmetry of the deformed Lagrangian. 
The Lagrangian $\mathcal{L}^{(0)}_{\mathcal{N} = 4}$ 
of $\mathcal{N}=4$ super Yang-Mills theory
is invariant 
under on-shell $\mathcal{N}=4$ supersymmetry, which is 
\begin{align}
\delta_{0}A_{\mu}&=i(\xi^{A}\sigma_{\mu}\bar{\Lambda}_{A}
+\bar{\xi}_{A}\bar{\sigma}_{\mu}\Lambda^{A}),
\notag\\
\delta_{0}\Lambda^{A}&=\sigma^{\mu\nu}\xi^{A}F_{\mu\nu}
+(\Sigma_{a})^{AB}\sigma^{\mu}\bar{\xi}_{B}D_{\mu}\varphi_{a}
-i(\Sigma_{ab})^{A}_{~B}\xi^{B}[\varphi_{a},\varphi_{b}],
\notag\\
\delta_{0}\bar{\Lambda}_{A}&=\bar{\sigma}^{\mu\nu}\bar{\xi}_{A}F_{\mu\nu}
+(\bar{\Sigma}_{a})_{AB}\bar{\sigma}^{\mu}\xi^{B}D_{\mu}\varphi_{a}
-i(\bar{\Sigma}_{ab})_{A}^{~B}\bar{\xi}_{B}[\varphi_{a},\varphi_{b}],
\notag\\
\delta_{0}\varphi_{a}&=i(\xi^{A}(\bar{\Sigma}_{a})_{AB}\Lambda^{B}
+\bar{\xi}_{A}(\Sigma_{a})^{AB}\bar{\Lambda}_{B}).
\end{align}
The deformed Lagrangian 
$\mathcal{L}^{(0)}_{\mathcal{N} = 4}+\mathcal{L}^{(1)}_{\mathrm{(S,A)}}
+\mathcal{L}^{(2)}_{\mathrm{(S,A)}}+\cdots$
is not invariant under this supersymmetry.
We explore deformation of supersymmetry under which the
deformed Lagrangian is invariant.
The deformed supersymmetry transformation $\delta$ 
can be expanded in the form 
$\delta=\delta_{0}+\delta_{1}+\cdots$, where $\delta_{n}$ is the
variation including of 
the $n$-th order power of $C$. 
The deformed supersymmetry 
$\delta_{n}$ is  determined recursively by solving 
the conditions \cite{ArItOh3, ItNa}
\begin{equation}
\delta_{1}\mathcal{L}^{(0)}_{\mathcal{N} = 4}
+\delta_{0}\mathcal{L}^{(1)}_{\mathrm{(S,A)}}=0, \quad
\delta_{2}\mathcal{L}^{(0)}_{\mathcal{N} = 4}
+\delta_{1}\mathcal{L}^{(1)}_{\mathrm{(S,A)}}
+\delta_{0}\mathcal{L}^{(2)}_{\mathrm{(S,A)}}=0,
\label{susy1}
\end{equation}
and so on.
However, we find that there is no solution of \eqref{susy1}
for generic $C$. 
In the first equation of (\ref{susy1}), a part of the variation
$\delta_0 \mathcal{L}^{(1)}_{\mathrm{(S,A)}}$ is canceled by 
deforming the supersymmetry transformation of $\Lambda^{A}$ as
\begin{equation}
\delta_{1}\Lambda^{A}=-i\varphi_{a}C_{\mu\nu a}\sigma^{\mu\nu}\xi^{A}.
\label{deform}
\end{equation}
Then, at the first order in $C$, we have
\begin{align}
\delta_{1}\mathcal{L}^{(0)}_{\mathcal{N} = 4}
+\delta_{0}\mathcal{L}^{(1)}_{\mathrm{(S,A)}}
&=
\frac{1}{kg_{\mathrm{YM}}^{2}}\mathrm{Tr}\Bigl[
-C^{(\alpha\beta)a}(\bar{\Sigma}_{a})_{AB}
\xi^{A}_{\beta}F_{\mu\nu}(\sigma^{\mu\nu}\Lambda^{B})_{\alpha}\notag\\
&\qquad{}-iC^{(\alpha\beta)a}(\bar{\Sigma}_{bc})_{A}^{~B}
(\bar{\Sigma}_{a})_{BC}\xi^{C}_{\alpha}
[\varphi_{b},\varphi_{c}]\Lambda^{A}_{\beta}
\Bigr]
\notag\\
&\quad{}+\frac{1}{kg_{\mathrm{YM}}^{2}}\mathrm{Tr}\Bigl[
-F_{\mu\nu}
C^{\mu\nu a}\bar{\xi}_{A}(\Sigma_{a})^{AB}\bar{\Lambda}_{B}\notag\\
&\qquad{}+C^{(\alpha\beta)a}\varphi_{b}(\bar{\Sigma}_{b})_{BA}(\Sigma_{a})^{AC}
(\sigma^{\mu}\bar{\xi}_{C})_{\alpha}D_{\mu}\Lambda^{B}_{\beta}\Bigr].
\label{ext1}
\end{align}
In order that the supersymmetry variation \eqref{ext1} vanishes, 
we have to require
\begin{equation}
\varepsilon_{ABCD}C^{(\alpha\beta)[BC]}\xi^{D}_{\beta}=0,\quad
C^{(\alpha\beta)[AB]}\bar{\xi}_{\dot{\alpha}B}=0,\label{sp1}
\end{equation}
which have only a trivial solution $\xi=\bar{\xi}=0$ for generic $C$.
The variation of the second order in $C$ also vanishes by the same
condition without introducing $\delta_2$. 
For special $C$ such that (\ref{sp1}) have nontrivial solution, 
the theory is invariant under the deformed supersymmetry
$\delta=\delta_0+\delta_1$ at the second order in $C$.
Although we do not fully classify the unbroken supersymmetries in
this paper,
we illustrate 
the number of deformed supersymmetry in the case where only 
$C^{(\alpha\beta)[12]}$ and $C^{(\alpha\beta)[34]}$ are nonzero.
{}From \eqref{sp1} the number of unbroken supersymmetry depends on 
the rank of $C^{(\alpha\beta)[12]}$ and $C^{(\alpha\beta)[34]}$.
We summarize the number of unbroken
supersymmetries in table \ref{n4susy},
where ${\cal N}=(p/2,q/2)$ denotes
supersymmetry with $p$ chiral and $q$ anti-chiral supercharges. 
\begin{table}[htbp]
\begin{center}
\begin{tabular}{|c|c|l|l|l|}
\hline
\multicolumn{2}{|c|}{ } & 
\multicolumn{3}{|c|}{rank of $C^{(\alpha\beta)[12]}$}\\
\cline{3-3}\cline{4-4}\cline{5-5}\multicolumn{2}{|c|}{ } & 
\multicolumn{1}{|c|}{0} & \multicolumn{1}{|c|}{1} & \multicolumn{1}{|c|}{2} \\
\hline
  & 0 & $\mathcal{N}=(2,2)$ & $\mathcal{N}=(3/2,1)$ & $\mathcal{N}=(1,1)$ \\
\cline{2-5}
rank of $C^{(\alpha\beta)[34]}$ & 1 & $\mathcal{N}=(3/2,1)$ & 
$\mathcal{N}=(1,0)$ & $\mathcal{N}=(1/2,0)$ \\
\cline{2-5}
 & 2 & $\mathcal{N}=(1,1)$ & $\mathcal{N}=(1/2,0)$ & $\mathcal{N}=(0,0)$ \\
\hline
\end{tabular}
\end{center}
\caption[tablesusy1]{The number of unbroken supersymmetry in 
$\mathcal{N}=4$ SYM with (S,A)-type deformation 
in the case where only 
$C^{(\alpha\beta)[12]}$ and $C^{(\alpha\beta)[34]}$ are nonzero.}
\label{n4susy}
\end{table}

\subsubsection{Deformed scalar potential}
In the case of non(anti)commutative $\mathcal{N}=4$
super Yang-Mills theory, fuzzy sphere configuration
with the constant $U(1)$ gauge field background is found \cite{Im, ItKoSa}.
In the deformed Lagrangian (\ref{N4sa1})-(\ref{N4sa2}), 
the scalar potential receives
also corrections from the R-R background. We investigate how
classical vacua configuration is deformed.

The scalar potential reads
\begin{align}
V(\varphi)&=
-\frac{1}{kg_{\mathrm{YM}}^{2}}\mathrm{Tr}\biggl[
\frac{1}{4}[\varphi_{a},\varphi_{b}]^{2}
+\frac{1}{2}(C^{\mu\nu a}\varphi_{a})^{2}
\biggr].
\end{align}
The stationary condition becomes
\begin{equation}
-\frac{\partial V(\varphi)}{\partial \varphi_{a}}=
\Bigl[\varphi_{b},[\varphi_{a},\varphi_{b}]\Bigr]
+C_{\mu\nu a}C^{\mu\nu b}\varphi_{b}
=0.\label{fseq2}
\end{equation}
We explore the solution with the fuzzy sphere ansatz such as
\begin{equation}
[\varphi_{a},\varphi_{b}]=if_{abc}\varphi_{c},
\label{fs2}
\end{equation}
where $f_{abc}$ is a constant antisymmetric tensor.
If we plug \eqref{fs2} into \eqref{fseq2}, we obtain
\begin{equation}
f_{abc}f_{bcd}=
C_{\mu\nu a}C^{\mu\nu}_{~~d}. \label{Killing}
\end{equation}
Hence \eqref{fs2} is a solution of \eqref{fseq2} if 
\eqref{Killing} is satisfied. 
We regard
$C^{\mu\nu}_{~~a}$ as a $6\times 6$ matrix
of which rows and columns are specified by $\mu\nu$ and $a$ respectively. 
One can find the rank of $C^{\mu\nu}_{~~a}$ is three 
due to the self-dual condition.
The rank of $C_{\mu\nu a}C^{\mu\nu}_{~~d}$ is also three. 
Then we can take the basis such that the upper-left $3\times 3$
submatrix of $C_{\mu\nu a}C^{\mu\nu}_{~~d}$  is only nonzero. 
The 
solution becomes
\begin{align}
[\varphi_{a},\varphi_{b}]&=i f_{abc}\varphi_{c} 
\quad\mbox{ for }a,b,c=5,6,7,
\label{fs}\\
[\varphi_{a},\varphi_{b}]&=0 ~~~~~~~\quad\mbox{ otherwise},
\end{align}
where 
$f_{abc}$ is totally antisymmetric tensor.
After the appropriate rescaling of $\varphi_{a}$, 
\eqref{fs} becomes 
the $SU(2)$ algebra\footnote{We assume that $f_{abc}$ are real.}.
Therefore \eqref{fs}
gives the fuzzy $\mathcal{S}^{2}$ solution.
We note that this fuzzy sphere configuration arises without the constant 
$U(1)$ gauge field strength background,
which is different from non(anti)commutative case 
\cite{Im, ItKoSa}.

\subsection{(A,S)-type deformation}
\subsubsection{Lagrangian}
In the (A,S)-type background, nonzero amplitudes with one
graviphoton vertex operator are given by
$\langle \! \langle V_{H_{\varphi \varphi}} V_{\varphi} V_{\mathcal{F}} 
\rangle \! \rangle$ and
$\langle \! \langle V_{\overline{\Lambda}} V_{\overline{\Lambda}} 
V_{\mathcal{F}} 
\rangle \! \rangle$,
which are evaluated as
\begin{eqnarray}
& & \langle \! \langle V^{(0)}_{H_{\varphi \varphi}} (p_1) V_{\varphi}^{(-1)} 
(p_2) V^{(-1/2,-1/2)}_{\mathcal{F}} \rangle \! 
 \rangle \nonumber \\
& & =  - \frac{1}{\sqrt{2}} \frac{ \pi i}{k g^2_
{\mathrm{YM}}} \mathrm{Tr} \left[
(\overline{\Sigma}^a \Sigma^b \overline{\Sigma}^c)_{AB} H_{ab} (p_1) 
\varphi_c (p_2) 
\right] (2 \pi \alpha')^{\frac{1}{2}} \mathcal{F}^{[\alpha \beta](AB)} 
\varepsilon_{\alpha \beta},
\end{eqnarray}
and
\begin{eqnarray}
& & \langle \! \langle V^{(-1/2)}_{\overline{\Lambda}} (p_1) 
V^{(-1/2)}_{\overline{\Lambda}} (p_2) V^{(-1/2,-1/2)}_{\mathcal{F}} \rangle
 \! \rangle \nonumber \\
& & = \frac{4 \pi i}{k g^2_{\mathrm{YM}}} \mathrm{Tr} \left[
\overline{\Lambda}_{\dot{\alpha} A} (p_1) 
\overline{\Lambda}^{\dot{\alpha}}_{\ B} (p_2) 
\right] (2 \pi \alpha')^{\frac{1}{2}} \mathcal{F}^{[\alpha \beta] (AB)} 
\varepsilon_{\alpha \beta}. 
\end{eqnarray}
After including the symmetric factor $1/2!$ for the second amplitude, 
we find 
that new interaction terms induced by the (A,S)-type background are
\begin{eqnarray}
- \frac{1}{\sqrt{2}} 
\frac{1}{k g^2_{\mathrm{YM}}} \mathrm{Tr} \left[
(\overline{\Sigma}^a \Sigma^b \overline{\Sigma}^c)_{AB} H_{ab} \varphi_c  
\right] C^{(AB)} 
+ \frac{2}{k g^2_{\mathrm{YM}}} \mathrm{Tr} \left[
\overline{\Lambda}_{\dot{\alpha} A}  
\overline{\Lambda}^{\dot{\alpha}}_{\ B}  
\right] C^{ (AB)},  
\end{eqnarray}
where
\begin{eqnarray}
C^{(AB)} \equiv - \pi i (2 \pi \alpha')^{\frac{1}{2}} 
 \mathcal{F}^{[\alpha \beta] (AB)} \varepsilon_{\alpha \beta}.
\end{eqnarray}
After integrating out the auxiliary fields, the deformed Lagrangian
is written as
${\cal L}^{(0)}_{\mathcal{N} = 4}+\mathcal{L}_{\mathrm{(A,S)}}^{(1)}
+\mathcal{L}_{\mathrm{(A,S)}}^{(2)}$, 
where
\begin{eqnarray}
\mathcal{L}_{\mathrm{(A,S)}}^{(1)}
 &=& \frac{1}{k g^2_{\mathrm{YM}}} 
 \mathrm{Tr} \left[
 (\overline{\Sigma}^a \Sigma^b \overline{\Sigma}^c)_{AB} \varphi_a 
\varphi_b \varphi_c  
\right] C^{(AB)} + \frac{2}{k g^2_{\mathrm{YM}}} \mathrm{Tr} \left[
\overline{\Lambda}_{\dot{\alpha} A}  
\overline{\Lambda}^{\dot{\alpha}}_{B}  
\right] C^{(AB)}, 
\label{N4as1}\\
\mathcal{L}_{\mathrm{(A,S)}}^{(2)}
&=& \frac{1}{4} \frac{1}{k g^2_{\mathrm{YM}}} \mathrm{Tr}
\left[ (\overline{\Sigma}^a \Sigma^b \overline{\Sigma}^c)_{AB}
(\overline{\Sigma}^a \Sigma^b \overline{\Sigma}^d)_{CD} \varphi_c
\varphi_d 
\right]
C^{(AB)} C^{(CD)}. \label{N4as2}
\end{eqnarray}
Here ${\cal L}^{(2)}_{\mathrm{(A,S)}}$
arises by integration over the auxiliary fields. 
In contrast to 
the (S,A)-type deformation, there are no other nonzero 
open string disk amplitudes at $O (C^2)$ in the case of (A,S)-type background. 
Therefore the $O(C^2)$ term is exact although there might exist
higher order deformed terms.

\subsubsection{Deformed supersymmetry}
We study 
supersymmetry of the deformed Lagrangian. 
As in the case of (S,A)-type deformation,
we expand supersymmetry transformation as 
$\delta=\delta_0+\delta_1+\cdots$.
Then the variation of the deformed Lagrangian at the first order in 
$C$ is
\begin{align}
\delta_{1}{\cal L}^{(0)}_{\mathcal{N} = 4}
+\delta_{0}\mathcal{L}^{(1)}_{\mathrm{(A,S)}}
&=
\frac{1}{k g^2_{\mathrm{YM}}}\mathrm{Tr}\Bigl[
6iC^{(AB)}(\bar{\Sigma}^{ab})_{B}^{~C}\varepsilon_{ACDE}\xi^{E}\Lambda^{D}
-4C^{(AB)}\bar{\xi}_{A}\bar{\sigma}^{\mu\nu}\bar{\Lambda}_{B}F_{\mu\nu}
\notag\\
&\qquad\qquad{}+iC^{(AB)}(\bar{\Sigma}^{ab})_{B}^{~C}
(2\bar{\xi}_{C}\bar{\Lambda}_{A}-6\bar{\xi}_{A}\bar{\Lambda}_{C})
[\varphi_{a},\varphi_{b}]
\Bigr],\label{ext2}
\end{align}
where we have deformed the supersymmetry transformation of $\Lambda^{A}$ as
\begin{equation}
\delta_{1}\Lambda^{A}=-4iC^{(AB)}(\bar{\Sigma}^{a})_{BC}\xi^{C}\varphi_{a}.
\end{equation}
The supersymmetry variation \eqref{ext2} vanishes if $C^{(AB)}$ satisfies
\begin{align}
&C^{(AB)}(\bar{\Sigma}^{ab})_{B}^{~C}\varepsilon_{ACDE}\xi^{E}=0,
\notag\\
&C^{(AB)}\bar{\xi}_{B}=0,\quad 
C^{(AB)}(\bar{\Sigma}^{ab})_{B}^{~C}\bar{\xi}_{C}=0.
\label{sp2}
\end{align}
The variation of second order in $C^{(AB)}$ also vanishes 
by the same condition without introducing $\delta_2$. 
If the rank of $C^{(AB)}$ is one, we have one nonzero $\xi^{A}$ and no nonzero
$\bar{\xi}_{A}$ as the solution of \eqref{sp2}. 
Then the supersymmetry is broken to $\mathcal{N}=(1/2,0)$.
If the rank of $C^{(AB)}$ is more than one, 
all supersymmetries are broken.

\subsubsection{Deformed scalar potential}
In the case of (A,S)-type deformation,
the  potential for the adjoint scalar field is
\begin{eqnarray}
- V (\varphi) &=& \frac{1}{4}  [\varphi_a, \varphi_b]^2 +
 (\overline{\Sigma}^a \Sigma^b \overline{\Sigma}^c)_{AB} \varphi_a
 \varphi_b \varphi_c C^{(AB)}
\nonumber\\
&&+ \frac{1}{4}  (\overline{\Sigma}^a \Sigma^b \overline{\Sigma}^c)_{AB}
 (\overline{\Sigma}^a \Sigma^b \overline{\Sigma}^d)_{CD}
\varphi_c \varphi_d C^{(AB)} C^{(CD)} . 
\end{eqnarray} 
The stationary condition is 
\begin{eqnarray}
- \frac{\partial V}{\partial \varphi_a} &=&
  [\varphi_b, [\varphi_a, \varphi_b]]
+ \frac{3}{2} (\overline{\Sigma}^a \Sigma^b \overline{\Sigma}^c)_{AB} C^{(AB)} 
[\varphi_b, \varphi_c] \nonumber\\
&&+ \frac{1}{2} (\overline{\Sigma}^a \Sigma^b
\overline{\Sigma}^c)_{AB}
(\overline{\Sigma}^b \Sigma^c \overline{\Sigma}^d)_{CD} C^{(AB)}
C^{(CD)} 
\varphi_d = 0.
\label{stat}
\end{eqnarray}
This equation has a fuzzy sphere solution. Let us assume that 
 $\varphi_a$ satisfies the commutation relation
\begin{eqnarray}
[\varphi_a, 
\varphi_b] = i \alpha f_{abc} \varphi_c \label{ansatz1},
\label{eq:commrel1}
\end{eqnarray}
where 
$f^{abc} = (\overline{\Sigma}^a \Sigma^b \overline{\Sigma}^c)_{AB} 
C^{(AB)}$.
The constant $\alpha$ is fixed by the equation
\begin{eqnarray}
\left( \alpha^2 - \frac{3}{2} i \alpha - \frac{1}{2} \right) f_{abc} 
 f_{bcd} \varphi_d = 0,
\label{eq:fuzzyas1}
\end{eqnarray}
which are obtained by the substitution of (\ref{eq:commrel1}) into 
the stationary condition (\ref{stat}).
The equation (\ref{eq:fuzzyas1}) admits nonzero
solutions. Therefore 
we can formally obtain the nontrivial fuzzy sphere solutions. 
However, $f_{abc}$ is subjected by the (imaginary) self-dual condition
\begin{equation}
f_{abc}=\frac{i}{3!}\varepsilon_{abcdef}f_{def}.
\end{equation}
For instance, if $f_{5,6,7}$ is real, $f_{8,9,10}$ is imaginary.
We should consider the fuzzy sphere configuration in the complexified
space of the scalar fields.

\section{Non-abelian Chern-Simons term \label{NACS}}
In this section,
we will check that the new bosonic interaction terms arising from the 
(S,A) and (A,S)-type backgrounds are consistent with the non-abelian 
Chern-Simons term in the D-brane effective action \cite{My}.
The Chern-Simons term is written as
\begin{eqnarray}
S_{CS} 
&=& \frac{\mu_3}{k} \mathrm{STr} \int_{\mathcal{M}_4}
\! \sum_n P [ e^{i \lambda \mathrm{i}_{\varphi}^{2} }
 \lambda^{\frac{1}{2}} \mathcal{A}^{(n)} ] e^{\lambda F}.
\end{eqnarray}
Here $\lambda = 2 \pi \alpha'$, $\mathcal{A}^{(n)}$ is an $n$-form R-R potential, 
$\mu_3 = \frac{1}{\lambda^2 g^2_{\mathrm{YM}}}$ is the R-R charge of
a D3-brane. 
The integral is performed over the 
four-dimensional D3-brane worldvolume $\mathcal{M}_4$. 
$F = \frac{1}{2 !} F_{\mu \nu} d x^{\mu} \wedge d x^{\nu}$ is 
a $U(N)$ gauge field strength which lives in the D3-brane worldvolume 
and $\varphi_a$ is $U(N)$ adjoint scalar fields. 
The symbol $P$ denotes the pull-back 
of ten-dimensional fields and $\mathrm{i}_{\varphi}$ is 
the interior product by $\varphi^a$.
STr is a symmetric trace of $U (N)$ gauge group. 
In the following, we will take a static gauge in which the 
four-dimensional part in ten-dimensional space-time is identified with 
the worldvolume direction. 

\subsection{(S,A)-type deformation}
For the (S,A)-type background, 
there exists the R-R 3-form and 
its dual 7-form field strength with the index structure
\begin{eqnarray}
\mathcal{F}_{\mu \nu a} &=& \partial_{[\mu} \mathcal{A}_{\nu] a} + \partial_a 
\mathcal{A}_{\mu \nu}, \nonumber \\
\mathcal{F}_{\mu \nu abcde} &=& \partial_{[\mu} \mathcal{A}_{\nu] abcde} + 
\partial_{(a} \mathcal{A}_{bcde) \mu \nu}, \label{sa-indices}
\end{eqnarray}
where $\mu, \nu = 1, \ldots, 4$ are worldvolume directions and 
$a, b, \ldots, e = 5, \ldots, 10$ are six-dimensional 
directions transverse to the D-brane worldvolume.

First, we calculate 
contributions from the 3-form field strength 
with the 2-form potentials
$\mathcal{A}_{\mu \nu}, \mathcal{A}_{\mu a}$. 
The Chern-Simons term is 
\begin{eqnarray}
\left. \frac{\mu_3}{k} \mathrm{STr} \int_{\mathcal{M}_4}
\! P [ e^{i \lambda \mathrm{i}_{\varphi}^{2}}
\lambda^{\frac{1}{2}} \mathcal{A}^{(2)} ] e^{\lambda F} \right|_{\mathrm{(S,A)}} 
= \frac{\mu_3}{k} \mathrm{STr} \frac{\lambda}{4} \int_{\mathcal{M}_4}
\! P [\mathcal{A}^{(2)}]_{\mu \nu} F_{\rho \sigma} \varepsilon^{\mu \nu \rho \sigma}
d^4 x  .
\end{eqnarray}
Here $|_{\mathrm{(S,A)}}$ means the restriction of the R-R indices to 
(S,A)-type deformation (\ref{sa-indices}). 
The pull-back is given by
\begin{eqnarray}
P [\mathcal{A}^{(2)}]_{\mu \nu} = \mathcal{A}_{MN} \frac{\partial X^M}{\partial x^{\mu}} 
\frac{\partial X^N}{\partial x^{\nu}} = \mathcal{A}_{\mu \nu} + 2 \lambda \mathcal{A}_{\mu a} 
D_{\nu} \varphi_a.
\end{eqnarray}
Here $X^M \ (M = (\mu,a) = 1, \ldots, 10)$ are
ten-dimensional space-time coordinates 
where 
$X^a$ are identified with adjoint 
scalar fields in $\mathcal{N} = 4$ vector multiplet through $X^a = 
\lambda \varphi_a$. Note that the pull-back is covariantized with respect to 
$U(N)$ gauge group. The potential has to be expanded by the fluctuation 
$\varphi_a$ 
such that
\begin{eqnarray}
& & \mathcal{A}_{\mu \nu} = \mathcal{A}^{(0)}_{\mu \nu} + \lambda \varphi_c \partial_c 
\mathcal{A}_{\mu \nu}^{(0)}, \nonumber \\
& & \mathcal{A}_{\mu a} = \mathcal{A}_{\mu a}^{(0)}.
\end{eqnarray}
Here $\mathcal{A}^{(0)}_{\mu \nu}, \mathcal{A}^{(0)}_{\mu a}$ are 2-form potentials evaluated 
at $\varphi_a = 0$. In the following we omit the 
superscript $(0)$.
After using Bianchi identity $\varepsilon^{\mu\nu\rho\sigma} D_{\nu} 
F_{\rho \sigma} = 0$ and partial integrations, we find
\begin{eqnarray}
 \frac{\mu_3}{k} \mathrm{STr} \left. \int_{\mathcal{M}_4} \! P [e^{i \lambda \mathrm{i}_{\varphi}^2 } 
\lambda^{\frac{1}{2}} \mathcal{A}^{(2)}
] e^{\lambda F} \right|_{\mathrm{(S,A)}} 
  = \frac{1}{2 k g^2_{\mathrm{YM}}} \int_{\mathcal{M}_4} \!\!\! d^4 x \  \mathrm{Tr} 
\left[ \varphi_a F_{\mu \nu} \right] (2 \pi \alpha')^{\frac{1}{2}} \mathcal{F}^{\mu \nu a}.
\end{eqnarray}
By identifying $(2 \pi \alpha')^{\frac{1}{2}}\mathcal{F}^{\mu \nu a} 
= 2 i C^{\mu \nu a}$, this Chern-Simons term precisely agrees with the 
$O(C)$ part of the (S,A)-deformation term 
(\ref{N4sa1}).

Next, we calculate contributions from the 7-form part, which take the form
\begin{eqnarray}
& & \frac{\mu_3}{k} \mathrm{STr} \int_{\mathcal{M}_4} \! \! 
P [e^{i \lambda \mathrm{i}_{\varphi}^{2}} 
\lambda^{\frac{1}{2}} \mathcal{A}^{(6)}] e^{\lambda F} \nonumber \\
& & \qquad = \frac{\mu_3}{k} \mathrm{STr} \int_{\mathcal{M}_4} \left[
i \lambda^{\frac{3}{2}} P [\mathrm{i}_{\varphi}^2 \mathcal{A}^{(6)}] - \frac{1}{2} \lambda^{\frac{7}{2}}
P [(\mathrm{i}_{\varphi}^2)^2 \mathcal{A}^{(6)}] \wedge F
- \frac{i}{2 \cdot 3!} \lambda^{\frac{11}{2}} P [(\mathrm{i}_{\varphi}^2)^3 \mathcal{A}^{(6)}]
\wedge F \wedge F \right], \nonumber \\
\label{eq:cssa1}
\end{eqnarray}
where $\mathcal{A}^{(6)}$ takes the form either $\mathcal{A}_{\mu abcde}$ or
$\mathcal{A}_{\mu \nu abcd}$. 
After evaluating STr, pull-back, and expansion in fluctuation, we find
that (\ref{eq:cssa1}) becomes
\begin{eqnarray}
&&
\frac{\lambda^{\frac{3}{2}}}{k g^2_{\mathrm{YM}}} 
\int_{\mathcal{M}_4} \! \! d^4 x \ \varepsilon^{\mu \nu \rho \sigma} 
\mathrm{STr} \left[\frac{i}{4} \mathcal{A}_{abcd \mu \nu} \varphi_b \varphi_a 
D_{\rho} \varphi_c D_{\sigma} \varphi_d - \frac{1}{8} \mathcal{A}_{abcd \mu \nu} 
\varphi_d \varphi_c \varphi_b \varphi_a F_{\rho \sigma} 
\right] \nonumber \\
& & \quad + \frac{\lambda^{\frac{5}{2}}}{k g^2_{\mathrm{YM}}} \int_{\mathcal{M}_4} \! \! d^4 x \ \varepsilon^{\mu \nu \rho \sigma} 
\mathrm{STr} \left[ \frac{i}{6} \mathcal{A}_{abcde \mu} \varphi_b 
\varphi_a D_{\nu} \varphi_c
D_{\rho} \varphi_d D_{\sigma} \varphi_e + \frac{i}{4} \partial_e \mathcal{A}_{abcd 
\mu \nu} \varphi_b \varphi_a \varphi_e D_{\rho} \varphi_c D_{\sigma} \varphi_d
\right. \nonumber \\
& & \qquad \qquad \qquad \qquad \qquad \qquad \left. - \frac{1}{8} \partial_e 
\mathcal{A}_{abcd \mu \nu} \varphi_d \varphi_c \varphi_b \varphi_a \varphi_e 
F_{\rho \sigma} - \frac{1}{4} \mathcal{A}_{abcde \mu} \varphi_d \varphi_c 
\varphi_b \varphi_a D_{\nu} \varphi_e F_{\rho \sigma} \right]. \nonumber \\
\end{eqnarray}
The $\lambda^{\frac{3}{2}}$ term vanishes by 
the partial integration.
The $\lambda^{\frac{5}{2}}$ 
term will not contribute to the deformation term in field theory limit in our 
scaling $\lambda^{\frac{1}{2}} \mathcal{F} = \mathrm{fixed}$. Thus 
we see that our open string calculation is consistent with effective 
action of D-brane in the presence of R-R background for the (S,A)-type 
deformation.

\subsection{(A,S)-type deformation}
For the (A,S)-type deformation, the R-R 3-form and
its dual 7-form field strength with index structure
are given by
\begin{eqnarray}
& & \mathcal{F}_{abc} = \partial_{(a} \mathcal{A}_{bc)}, \nonumber \\
& & \mathcal{F}_{\mu \nu \rho \sigma abc} = \partial_{[\mu} \mathcal{A}_{\nu \rho 
\sigma] abc} + \partial_{(a} \mathcal{A}_{bc) \mu \nu \rho \sigma}.
\end{eqnarray}
The Chern-Simons term corresponding to the R-R 2-form potential is
\begin{eqnarray}
\frac{\mu_3}{k} \mathrm{STr} \left. \int_{\mathcal{M}_4}  \ P [e^{i \lambda
	      \mathrm{i}_{\varphi}^{2} }
\lambda^{\frac{1}{2}} \mathcal{A}^{(2)} ] e^{\lambda F} \right|_{\mathrm{(A,S)}}
&=& \frac{\lambda^{\frac{5}{2}}}{4 k g^2_{\mathrm{YM}}} \mathrm{STr} 
\int_{\mathcal{M}_4} \!\! d^4 x \ \partial_c \mathcal{A}_{ab}
\varphi_c D_{\mu} \varphi_a D_{\nu} \varphi_b
F_{\rho \sigma} \varepsilon^{\mu \nu \rho \sigma}
\nonumber \\
& & + \frac{i\lambda^{\frac{5}{2}}}{8 k g^2_{\mathrm{YM}}} \mathrm{STr} \int_{\mathcal{M}_4} \!\! 
d^4 x \ \partial_c \mathcal{A}_{ab} \varphi_b \varphi_a  \varphi_c F_{\mu \nu} F_{\rho 
\sigma} \varepsilon^{\mu \nu \rho \sigma}. \nonumber \\
\end{eqnarray}
After evaluating STr and performing partial integrations, we find 
that this becomes
\begin{eqnarray}
&&
\frac{1}{12 k g^2_{\mathrm{YM}}} 
\int_{\mathcal{M}_4} \! d^4 x \ 
\mathrm{Tr} \left[ \frac{}{}  \varphi_a
D_{\mu} \varphi_b D_{\nu} \varphi_c F_{\rho \sigma} \varepsilon^{\mu \nu 
\rho \sigma} \right] (2 \pi \alpha')^{\frac{5}{2}} \mathcal{F}_{abc}
\nonumber \\
& & \qquad - \frac{i}{24 k g^2_{\mathrm{YM}}} 
\int_{\mathcal{M}_4} \! d^4 x \ \mathrm{Tr} \left[ \frac{}{}
\varphi_a \varphi_b \varphi_c F_{\mu \nu} F_{\rho \sigma} 
\varepsilon^{\mu \nu \rho \sigma} \right] (2 \pi \alpha')^{\frac{5}{2}}
\mathcal{F}_{abc}. 
\end{eqnarray}
Those terms vanish in the zero-slope limit 
$\alpha' \to 0$ with fixed 
$\lambda^{\frac{1}{2}} \mathcal{F}$.
On the other hand, the 7-form part is calculated by the same way as 
\begin{eqnarray}
\left. \frac{\mu_3}{k} \mathrm{STr} \int_{\mathcal{M}_4} \! \! P [e^{i \lambda
 \mathrm{i}_{\varphi}^2}
\lambda^{\frac{1}{2}} \mathcal{A}^{(6)}] e^{\lambda F} \right|_{\mathrm{(A,S)}} 
= - \frac{i}{3 \cdot 4! } \frac{1}{k g^2_{\mathrm{YM}}}
\int_{\mathcal{M}_4} \!\! d^4 x \ 
\mathrm{Tr} \left[ \frac{}{} \varphi_a \varphi_b \varphi_c \right]
(2 \pi \alpha')^{\frac{1}{2}} \tilde{\mathcal{F}}_{abc}. \nonumber \\
\end{eqnarray}
Here we have defined 
\begin{eqnarray}
\tilde{\mathcal{F}}_{abc} \equiv \mathcal{F}_{abc \mu \nu \rho \sigma}
\varepsilon^{\mu \nu \rho \sigma}.
\end{eqnarray}
This term precisely agrees with the (A,S)-type deformation term 
(\ref{N4as1}) at linear order 
in deformation parameter with the identification $- \frac{i}{3 \cdot 4!}
(2 \pi \alpha')^{\frac{1}{2}} \tilde{\mathcal{F}}^{abc} 
= (\overline{\Sigma}^a \Sigma^b \overline{\Sigma}^c)_{AB} C^{(AB)}$.
Therefore the (A,S)-type deformation is related to the dual 7-form
R-R field strength.

\section{Deformed $\mathcal{N}=2$ super Yang-Mills theories 
\label{N2deformation}}
So far we have studied the deformation of ${\cal N}=4$ super Yang-Mills
theory in the R-R 3-form background.
In this section we study deformed $\mathcal{N}=2$ $U(N)$ super Yang-Mills
theory
in the (S,A) and (A,S)-type backgrounds.
To realize $\mathcal{N} = 2$ $U(N)$ supersymmetric gauge theory, 
we use $N$ fractional D3-branes located at the singularity of the
orbifold ${\bf C}^2/{\bf Z}_2$\cite{BeDiVeFrLeMa}.
Since the orbifold projection
restricts  $R$-symmetry group $SU(4)$ to $SU(2)$,
the internal spin fields $S_A$ become the doublet $S_i$ ($i=1,2$) 
of $SU(2)$.
The massless fields on the fractional D3-branes are 
gauge fields $A_\mu$,  Weyl fermions $\Lambda^i_\alpha$ and a complex
scalar $\varphi$, 
whose vertex operators are obtained by the orbifold projection and
are defined in \cite{ItSa}. 
The undeformed Lagrangian is given by
\begin{eqnarray}
\mathcal{L}^{(0)}_{\mathcal{N} = 2} &=&  \frac{1}{kg^2_{\mathrm{YM}}} 
\mathrm{Tr} \left[ - \frac{1}{4} F_{\mu \nu}
 F^{\mu \nu} - \frac{1}{4} F_{\mu \nu}
 \tilde{F}^{\mu \nu} - D_{\mu} \varphi D^{\mu} \bar{\varphi} - \frac{1}{2} 
 [\varphi, \bar{\varphi}]^2   \right. \nonumber \\
& & \qquad \qquad \qquad \left. - i \Lambda^{i \alpha} 
			  (\sigma^{\mu})_{\alpha \dot{\beta}} D_{\mu} 
			  \overline{\Lambda}_i^{\ \dot{\beta}}
-  \frac{i}{\sqrt{2}} \Lambda^i [\bar{\varphi}, \Lambda_i] +
\frac{i}{\sqrt{2}} \overline{\Lambda}_i [\varphi, \overline{\Lambda}^i] \right].
\end{eqnarray}
We introduce
the R-R vertex operator of the form
\begin{eqnarray}
V_{\mathcal{F}}^{(-1/2,-1/2)} (z, \bar{z})  \!=\!  (2 \pi \alpha') 
\mathcal{F}^{\alpha \beta i j}  \left[ S_{\alpha} (z) S^{(-)} (z) S_i (z) 
e^{- \frac{1}{2} \phi (z)} S_{\beta} (\bar{z}) S^{(-)} (\bar{z}) S_j 
(\bar{z}) e^{ - \frac{1}{2} \phi (\bar{z})} \right].
\end{eqnarray}
The R-R field strength can be decomposed into 
 $\mathcal{F}^{(\alpha \beta) 
(ij)}$,  $\mathcal{F}^{(\alpha \beta) [ij]}$,
 $\mathcal{F}^{[\alpha \beta] (ij) }$ and 
 $\mathcal{F}^{[\alpha \beta] [ij]}$,
which corresponds to the R-R 5-form, 3-form (7-form), 
3-form (7-form) and 1-form (9-form) field strength respectively. 
We calculate the deformed Lagrangian in the (S,A) and (A,S)-type
deformations with the scaling condition ${\cal F}\sim (\alpha')^{-1/2}$
as we did in the ${\cal N}=4$ case.

\subsection{(S,A)-type deformation}
The $\mathcal{N} = 2$ (S,A)-type deformation was studied in \cite{BiFrFuLe}.
The nonzero disk amplitudes which contain single 
$\mathcal{F}^{(\alpha \beta) [ij]}$, 
are $ \langle \! \langle V_{A} V_{\bar{\varphi}} 
V_{\mathcal{F}} \rangle \! \rangle $ and $ \langle \! \langle V_{H_{AA}} 
V_{\bar{\varphi}} V_{\mathcal{F}} \rangle \! \rangle $.
The first amplitude is evaluated as
\begin{eqnarray}
& & \langle \! \langle V^{(0)}_A (p_1) V^{(-1)}_{\bar{\varphi}} (p_2) 
V^{{(-1/2,-1/2)}}_{\mathcal{F}} \rangle \! \rangle 
\nonumber \\
& & = \frac{4 \sqrt{2} \pi}{k g^2_{\mathrm{YM}}} \mathrm{Tr}
\left[ (\sigma^{\mu \nu})_{\alpha \beta} i p_{1 \mu} A_{\nu} (p_1)
\bar{\varphi} (p_2) \right] (2 \pi \alpha')^{\frac{1}{2}} 
\mathcal{F}^{(\alpha \beta) [ij]} \varepsilon_{ij}.
\end{eqnarray}
Combining the result of the second amplitude, we get the 
interaction term
\begin{eqnarray}
 - (-i) \frac{2 \sqrt{2} \pi}{k g^2_{\mathrm{YM}}} 
\mathrm{Tr} \left[ \left( \partial_{[\mu} A_{\nu]} - \frac{i}{2}
H_{\mu \nu} \right) \bar{\varphi} (\sigma^{\mu \nu})_{\alpha \beta} \right]
\varepsilon_{ij} \mathcal{F}^{(\alpha \beta) [ij]} .
\end{eqnarray}
After integrating out the auxiliary fields, we find
\begin{eqnarray}
\mathcal{L}^{(1)}_{\mathrm{(S,A)}}
+\mathcal{L}^{(2)}_{\mathrm{(S,A)}}
 = \frac{1}{k g^2_{\mathrm{YM}} } \mathrm{Tr} 
\left[i F_{\mu \nu} \bar{\varphi} \tilde{C}^{\mu \nu}
+ \frac{1}{2} (\bar{\varphi} \tilde{C}^{\mu \nu})^2 \right],  \label{n2sa}
\end{eqnarray}
where we have defined $\tilde{C}^{\mu \nu} \equiv  2 \sqrt{2} \pi i (\sigma^{\mu \nu})_{\alpha \beta} 
\varepsilon_{ij} \mathcal{F}^{(\alpha \beta) [ij]}$. 
Since at order $C^2$ there are no other disk amplitudes which 
 contribute to the Lagrangian, 
the deformed Lagrangian is exact up to higher order
corrections in $C$.

The deformation term \eqref{n2sa} can be also obtained by the reduction from 
$\mathcal{N}=4$ to $\mathcal{N}=2$ by 
the $\mathbf{Z}_{2}$ orbifold projection,
which is given by
\begin{equation}
\Lambda^{A}_{\alpha}=0\mbox{ for }A=3,4,\quad
\varphi_{a}=0\mbox{ for }a=7,8,9,10,
\end{equation}
and only $C^{(\alpha\beta)[12]}$ and $C^{(\alpha\beta)[34]}$ are nonzero
\cite{BiFrFuLe}.
Under the reduction, the deformation term becomes
\begin{equation}
\mathcal{L}^{(1)}_{\mathrm{(S,A)}}
+\mathcal{L}^{(2)}_{\mathrm{(S,A)}}=
\frac{1}{k g^2_{\mathrm{YM}} } \mathrm{Tr}\biggl[
i(\tilde{C}^{\mu\nu}\bar{\varphi}+\bar{C}^{\mu\nu}\varphi)
F_{\mu\nu}
-\frac{1}{\sqrt{2}}\bar{C}^{\mu\nu}
\Lambda^{i}\sigma_{\mu\nu}\Lambda_{i}
+\frac{1}{2}
(\tilde{C}^{\mu\nu}\bar{\varphi}+\bar{C}^{\mu\nu}\varphi)^{2}
\biggr],\label{n2sa2}
\end{equation}
where $\tilde{C}^{\mu\nu}$ and $\bar{C}^{\mu\nu}$ are
defined as
\begin{equation}
\tilde{C}^{\mu\nu}=2\sqrt{2}iC^{\mu\nu [12]}, \quad 
\bar{C}^{\mu\nu}=-2\sqrt{2}iC^{\mu\nu [34]},
\end{equation}
and we have used
\begin{equation}
\varphi=\frac{1}{\sqrt{2}}(\varphi_{5} - i \varphi_{6}),\quad
\bar{\varphi}=\frac{1}{\sqrt{2}}(\varphi_{5} + i \varphi_{6}).
\end{equation}
In the case of $\bar{C}^{\mu\nu}=0$,
\eqref{n2sa2} is reduced to
\eqref{n2sa}.
The deformation parameter 
$\bar{C}^{\mu\nu}$
is referred as 
the graviphoton-like vertex operator 
in \cite{BiFrFuLe}.

We examine the deformed supersymmetry of the 
Lagrangian ${\cal L}^{(0)}_{\mathcal{N} = 2}+{\cal
L}^{(1)}_{\mathrm{(S,A)}}
+{\cal L}^{(2)}_{\mathrm{(S,A)}}$. 
The deformed supersymmetry transformation is obtained from the 
${\bf Z}_2$ projection in ${\cal N}=4$ theory, which is given by
\begin{align}
\delta A_{\mu}&=i(\xi^{i}\sigma_{\mu}\bar{\Lambda}_{i}
+\bar{\xi}_{i}\bar{\sigma}_{\mu}\Lambda^{i}),
\notag\\
\delta \Lambda^{i}&=\sigma^{\mu\nu}\xi^{i}\bigl(F_{\mu\nu}
-i(\tilde{C}^{\mu\nu}\bar{\varphi}+\bar{C}^{\mu\nu}\varphi)\bigr)
+\sqrt{2}i\sigma^{\mu}\bar{\xi}^{i}D_{\mu}\varphi
-i\xi^{i}[\varphi,\bar{\varphi}],
\notag\\
\delta \bar{\Lambda}_{i}&=\bar{\sigma}^{\mu\nu}\bar{\xi}_{i}F_{\mu\nu}
-\sqrt{2}i\bar{\sigma}^{\mu}\xi_{i}D_{\mu}\bar{\varphi}
+i\bar{\xi}_{i}[\varphi,\bar{\varphi}],
\notag\\
\delta\varphi&=\sqrt{2}\xi^{i}\Lambda_{i},
\notag\\
\delta\bar{\varphi}&=\sqrt{2}\bar{\xi}^{i}\bar{\Lambda}_{i}.
\label{N2sasusy}
\end{align}
The deformed Lagrangian is invariant under 
(\ref{N2sasusy})
if $\xi$ and $\bar{\xi}$ satisfy
\begin{align}
&\bar{C}^{(\alpha\beta)}\xi^{i}_{\beta}=0,\notag\\
&\bar{\xi}_{i}=0\ \mbox{or}\ \tilde{C}^{(\alpha\beta)}=0,
\end{align}
where $\tilde{C}^{(\alpha\beta)}=2\sqrt{2}iC^{(\alpha\beta)[12]}$, 
$\bar{C}^{(\alpha\beta)}=-2\sqrt{2}iC^{(\alpha\beta)[34]}$.
As in the $\mathcal{N}=4$ case, we can classify the unbroken 
supersymmetries, which are summarized in 
table \ref{n2susy}.
\begin{table}[htbp]
\begin{center}
\begin{tabular}{|c|c|l|l|l|}
\hline
\multicolumn{2}{|c|}{ } & 
\multicolumn{3}{|c|}{rank of $\tilde{C}^{(\alpha\beta)}$}\\
\cline{3-3}\cline{4-4}\cline{5-5}\multicolumn{2}{|c|}{ } & 
\multicolumn{1}{|c|}{0} & \multicolumn{1}{|c|}{1} & \multicolumn{1}{|c|}{2} \\
\hline
  & 0 & $\mathcal{N}=(1,1)$ & $\mathcal{N}=(1,0)$ & $\mathcal{N}=(1,0)$ \\
\cline{2-5}
rank of $\bar{C}^{(\alpha\beta)}$ & 1 & $\mathcal{N}=(1/2,1)$ & 
$\mathcal{N}=(1/2,0)$ & $\mathcal{N}=(1/2,0)$ \\
\cline{2-5}
 & 2 & $\mathcal{N}=(0,1)$ & $\mathcal{N}=(0,0)$ & $\mathcal{N}=(0,0)$ \\
\hline
\end{tabular}
\end{center}
\caption[tablesusy2]{The number of unbroken supersymmetry in 
$\mathcal{N}=2$ SYM with (S,A)-type deformation. }
\label{n2susy}
\end{table}

\subsection{(A,S)-type deformation}
Next we consider the (A,S)-type deformation of ${\cal N}=2$ super
Yang-Mills theory.
At the first order in $\mathcal{F}$, the 
nonzero amplitude is possible only for 
\begin{eqnarray}
& & \langle \! \langle V^{(-1/2)}_{\overline{\Lambda}} (p_1) 
V^{(-1/2)}_{\overline{\Lambda}} (p_2) V^{(-1/2,-1/2)}_{\mathcal{F}}
\rangle \! \rangle \nonumber \\
& & \qquad = \frac{4 \pi i}{k g^2_{\mathrm{YM}}} \mathrm{Tr} \left[
\overline{\Lambda}_{\dot{\alpha} i} (p_1) 
\overline{\Lambda}^{\dot{\alpha}}_{\ j} (p_2) 
\right] (2 \pi \alpha')^{\frac{1}{2}} \mathcal{F}^{[\alpha \beta] (ij)}
\varepsilon_{\alpha \beta}.
\end{eqnarray}
The interaction term is given by
\begin{eqnarray}
\mathcal{L}^{(1)}_{\mathrm{(A,S)}}
 =  \frac{1}{k g^2_{\mathrm{YM}}} \mathrm{Tr} \left[
\overline{\Lambda}_{\dot{\alpha} i} (x) 
\overline{\Lambda}^{\dot{\alpha}}_{\ j} (x) 
\right] C^{(ij)}. \label{n2as}
\end{eqnarray}
Here $C^{(ij)} \equiv - 2 \pi i (2 \pi \alpha')^{\frac{1}{2}} 
\mathcal{F}^{[\alpha \beta] (ij)} \varepsilon_{\alpha \beta}$.

As in the case of (S,A)-type deformation, We can obtain deformed 
Lagrangian from the ${\cal N}=4$ one by the reduction. 
The deformation parameter $C^{(AB)}$ 
takes the block diagonal form:
\begin{equation}
C^{(AB)}=\frac{1}{2}\begin{pmatrix} C^{(ij)} & 0 \\ 0 & C^{(\hat{i}\hat{j})} 
\end{pmatrix},\quad i,j=1,2,\quad \hat{i},\hat{j}=3,4.
\end{equation}
Then the deformation terms become
\begin{equation}
\mathcal{L}^{(1)}_{\mathrm{(A,S)}}
+\mathcal{L}^{(2)}_{\mathrm{(A,S)}}=
\frac{1}{k g^2_{\mathrm{YM}}}\mathrm{Tr}\Bigl[
C^{(ij)}\bar{\Lambda}_{\dot{\alpha}i}\bar{\Lambda}^{\dot{\alpha}}_{j}
-C^{(ij)}C_{(ij)}\bar{\varphi}^{2}
-C^{(\hat{i}\hat{j})}C_{(\hat{i}\hat{j})}\varphi^{2}
\Bigr]. \label{N4sa_reduction}
\end{equation}
We note that only the $O(C^2)$ terms 
in (\ref{N4sa_reduction}) 
are allowed to exist at this order due to
the charge conservation of vertex operators in the disk amplitudes,
which are given by 
$\langle \! \langle V_{\bar{\varphi}} V_{\bar{\varphi}} V_{\mathcal{F}}
V_{\mathcal{F}} \rangle \! \rangle $, $\langle \! \langle 
V_{\varphi} V_{\varphi} V_{\mathcal{\bar{F}}} V_{\bar{\mathcal{F}}} 
\rangle \! \rangle$.
Here $V_{\bar{\mathcal{F}}}$ is the closed string R-R vertex operator
corresponding to $C^{\hat{i}\hat{j}}$. 

The deformed Lagrangian is invariant under the supersymmetry transformation 
\begin{align}
\delta A_{\mu}&=i(\xi^{i}\sigma_{\mu}\bar{\Lambda}_{i}
+\bar{\xi}_{i}\bar{\sigma}_{\mu}\Lambda^{i}),
\notag\\
\delta \Lambda^{i}&=\sigma^{\mu\nu}\xi^{i}F_{\mu\nu}
+\sqrt{2}i\sigma^{\mu}\bar{\xi}^{i}D_{\mu}\varphi
-i\xi^{i}[\varphi,\bar{\varphi}]
-4\sqrt{2}\bar{\varphi}C^{(ij)}\xi_{j},
\notag\\
\delta \bar{\Lambda}_{i}&=\bar{\sigma}^{\mu\nu}\bar{\xi}_{i}F_{\mu\nu}
-\sqrt{2}i\bar{\sigma}^{\mu}\xi_{i}D_{\mu}\bar{\varphi}
+i\bar{\xi}_{i}[\varphi,\bar{\varphi}],
\notag\\
\delta\varphi&=\sqrt{2}\xi^{i}\Lambda_{i},
\notag\\
\delta\bar{\varphi}&=\sqrt{2}\bar{\xi}^{i}\bar{\Lambda}_{i},
\label{N2assusy}
\end{align}
if $\bar{\xi}$ satisfies
\begin{equation}
C^{(ij)}\bar{\xi}_{j}=0.
\end{equation}
Hence the theory has 
$\mathcal{N}=(1,0)$ supersymmetry in the generic case. 
But it is enhanced to $\mathcal{N}=(1,1/2)$ supersymmetry 
if the rank of $C^{(ij)}$ is one.

\subsection{Comments on the  reduction to $\mathcal{N} = 1$ theory}
We are able to discuss further reduction to deformed ${\cal N}=1$
theory from the orbifold 
${\bf R}^6/{\bf Z}_2\times {\bf Z}_2$ \cite{BeDiVeFeMa},
which can be done by restriction
$\varphi_a = 0 \ (a = 5, \ldots 10), \Lambda^{2,3,4} = 
\overline{\Lambda}^{2,3,4} = 0$ in ${\cal N}=4$ theory. 
The deformation parameter ${\cal F}^{\alpha\beta AB}$ remains nonzero for
$A=B=1$.
Therefore it is easy to see that (S,A)-type deformation with parameter
${\cal F}^{\alpha\beta [AB]}$ does not exist
in ${\cal N}=1$ theory.

On the other hand, the (A,S)-type deformation is still allowed in ${\cal N}=1$
theory.
In fact the reduction from ${\cal N}=4$ theory leads to
the interaction term
\begin{eqnarray}
\mathcal{L}_{\mathrm{(A,S)}} = \frac{1}{k g^2_{\mathrm{YM}}} \mathrm{Tr} \left[
\overline{\Lambda}_{\dot{\alpha}} \overline{\Lambda}^{\dot{\alpha}} 
\right] C,
\end{eqnarray}
where $C= 2 C^{(11)[\alpha \beta]} \varepsilon_{\alpha \beta}$.
This result is also consistent with direct computation of string amplitudes.

We note that it is possible to deform ${\cal N}=1$ super Yang-Mills
theory in the (S,S)-type background with the scaling condition
$(2 \pi \alpha')^{\frac{1}{2}}\mathcal{F} = \mathrm{fixed}$, 
where the scaling condition
$(2 \pi \alpha')^{\frac{3}{2}} \mathcal{F} = \mathrm{fixed}$ leads
to a non(anti)commutative deformation of superspace \cite{BiFrPeLe, ItSa,
ItKoSa}.  
We find, however, that there are no interaction terms in the 
zero slope limit from calculation of disk amplitudes and Chern-Simons
term. We conclude ${\cal N}=1$ super Yang-Mills theory is not deformed 
in the (S,S)-type background at least up to leading order in deformation 
parameter.

\section{Deformed Lagrangian in ${\cal N}=1$ superspace \label{superspace}}
Although the (S,A) and (A,S) type deformation is not realized as 
non(anti)commutative superspace deformation, it would be useful to rewrite 
the deformation Lagrangian in 
superfields in order to understand its geometrical structure.
In this section we explore a geometrical interpretation of 
the deformed super Yang-Mills theories in terms of ${\cal N}=1$
superspace.

\subsection{$\mathcal{N} = 4$ deformation}
The Lagrangian of $\mathcal{N} = 4$ super Yang-Mills theory
in $\mathcal{N} = 1$ superspace is given by
\begin{eqnarray}
{\cal L}^{\mathcal{N} = 4}
&=& \frac{1}{k g^2_{\mathrm{YM}}} \int \! d^2 \theta d^2 \bar{\theta} \ 
 \mathrm{Tr} \sum^3_{i=1} \left( \overline{\Phi}_i  e^{2V}  \Phi_i  
 e^{-2V}  \right) 
+ \frac{1}{16 k g^2_{\mathrm{YM}}}\mathrm{Tr}  \biggl[\int \! d^2 \theta \,
W^{\alpha}  W_{\alpha} + 
\!\int \! d^2 \bar{\theta} \,
\overline{W}_{\dot{\alpha}}  \overline{W}^{\dot{\alpha}} 
\biggr]
\nonumber \\
& & - \frac{\sqrt{2}}{3} \frac{1}{k g^2_{\mathrm{YM}}} \int \! d^2 \theta \ 
\mathrm{Tr} \varepsilon^{ijk} \left( \Phi_i  \Phi_j  \Phi_k \right)
+ \frac{\sqrt{2}}{3} \frac{1}{k g^2_{\mathrm{YM}}} \int \! d^2 \bar{\theta} \
\mathrm{Tr} \varepsilon^{ijk} \left( \overline{\Phi}_i  \overline{\Phi}_j 
	      \overline{\Phi}_k \right) . 
\end{eqnarray}
Here $\Phi_i,(\overline{\Phi}_i)$ $(i=1,2,3)$ are (anti-)chiral superfields,
$V$ a vector superfield, $W_{\alpha}, \overline{W}_{\dot{\alpha}}$
its super field strengths, which are written in terms of component fields
as
\begin{eqnarray}
& & \Phi_i = \phi_i (y) + \sqrt{2} \theta \psi_i (y) + \theta \theta F_i 
(y), \nonumber\\
& & \overline{\Phi}_i = \bar{\phi}_i (\bar{y}) + \sqrt{2} \bar{\theta}
 \bar{\psi} _i
(\bar{y})  + \bar{\theta} \bar{\theta} \bar{F}_i 
(\bar{y}), \nonumber\\
& & 2^{-1} W_{\alpha} = - i \lambda_{\alpha} + \left[ \delta_{\alpha} 
{}^{\beta} D - i (\sigma^{\mu \nu})_{\alpha} {}^{\beta} F_{\mu \nu} 
\right] \theta_{\beta} + \theta^2 (\sigma^{\mu})_{\alpha \dot{\alpha}} 
D_{\mu} \bar{\lambda}^{\dot{\alpha}}, \nonumber\\
& & 2^{-1} \overline{W}_{\dot{\alpha}} 
= - i \bar{\lambda}_{\dot{\alpha}} + \left[ \varepsilon_{\dot{\alpha}
\dot{\beta}}  D + i \varepsilon_{\dot{\alpha} \dot{\gamma}}
(\bar{\sigma}^{\mu \nu})^{\dot{\gamma}} {}_{\dot{\beta}} F_{\mu \nu} 
\right] \bar{\theta}^{\dot{\beta}} - \varepsilon_{\dot{\alpha} \dot{\beta}} 
\bar{\theta}^2 (\bar{\sigma}^{\mu})^{\dot{\beta} \alpha} D_{\mu} \lambda_{\alpha}.
\end{eqnarray}
We have followed the notation and convention in \cite{WeBa}.

Firstly we consider the (S,A)-type deformation.
We can show that the 
interaction terms (\ref{N4sa1}) and (\ref{N4sa2}) are regarded
as the deformation of D-terms and F-terms:
\begin{eqnarray}
& & \mathcal{L}_{\mathrm{(S,A)}}^{(1)} + \mathcal{L}^{(2)}_{\mathrm{(S,A)}}
\nonumber \\
& & =  
\frac{1}{2 k g^2_{\mathrm{YM}}} 
\int \! d^4 \theta \ \theta^2 \bar{\theta}^2 \mathrm{Tr} \Bigl[
\left( \overline{\Phi}_1 C^{(\alpha \beta) [12]}
+ \overline{\Phi}_2 C^{(\alpha \beta) [31]}
+ \overline{\Phi}_3 C^{(\alpha \beta) [14]} \right)
D_{\alpha} W_{\beta} \Bigr] \nonumber \\
& & - \frac{4}{k g^2_{\mathrm{YM}}} \int \! d^2 \theta \ \theta^2 
\mathrm{Tr} \Bigl[
D_{\alpha} \Phi_1 D_{\beta} \Phi_2 C^{(\alpha \beta) [14]} 
+ D_{\alpha} \Phi_2 D_{\beta} \Phi_3 C^{(\alpha \beta) [12]} 
+ D_{\alpha} \Phi_3 D_{\beta} \Phi_1 C^{(\alpha \beta) [13]} 
\Bigr] \nonumber \\
& & +
\frac{\sqrt{2}}{k g^2_{\mathrm{YM}}}  
\int \! d^2 \theta \ \theta^2 \mathrm{Tr} 
\Bigl[
\left( D_{\alpha} \Phi_1 W_{\beta} + \Phi_1 D_{\alpha} W_{\beta}
\right) C^{(\alpha \beta) [34]} + 
\left( D_{\alpha} \Phi_2 W_{\beta} + \Phi_2 D_{\alpha} W_{\beta}
\right) C^{(\alpha \beta) [24]} \nonumber \\
& & \qquad \qquad \qquad \qquad \qquad + \frac{}{}
\left( D_{\alpha} \Phi_3 W_{\beta} + \Phi_3 D_{\alpha} W_{\beta}
\right) C^{(\alpha \beta) [23]} \Bigr] \nonumber \\
& & + \frac{4}{k g^2_{\mathrm{YM}}}  \int \! d^2 \theta \ 
\theta^2 \mathrm{Tr} \Bigl[ 
( \Phi_1 C^{\mu \nu [34]} + \Phi_2 C^{\mu \nu [42]} 
+ \Phi_3 C^{\mu \nu [23]} )^2
\Bigr] \nonumber \\
& & + \frac{4}{kg^2_{\mathrm{YM}}} \int \! d^2 \bar{\theta} 
\ \bar{\theta}^2 \mathrm{Tr} \Bigl[ ( \overline{\Phi}_1 C^{\mu \nu [12]}
+ \overline{\Phi}_2 C^{\mu \nu [31]} + \overline{\Phi}_3 C^{\mu \nu [14]}
)^2 \Bigr].
\end{eqnarray}
Here we have used the relation
\begin{eqnarray}
\varphi_{(2 i - 1) + 4} = \frac{1}{\sqrt{2}} 
\left( \phi_i + \bar{\phi}_i \right),
\quad \varphi_{2i+4} = \frac{i}{\sqrt{2} } 
\left( \phi_i - \bar{\phi}_i \right).
\end{eqnarray}
It is natural to think that this complicated expression is simplified if 
one uses ${\cal N}=2$ superspace formalism as in \cite{LoMaNe}, which will
be discussed elsewhere.

Next, we study the (A,S)-type deformation.
In this case we have simple interpretation of the Lagrangian
in terms of deformation 
of gauge coupling constants and 
complex mass parameters, which are functions
on ${\cal N}=1$ superspace.
To see this, let us consider 
 generic mass deformation of the 
${\cal N}=4$ Lagrangian
\begin{eqnarray}
\mathcal{L}_m^{\mathcal{N} = 4} 
&=& \mathcal{L}^{\mathcal{N} = 4} +
 \frac{1}{2 k g^2_{\mathrm{YM}}} \int \! d^2 \theta \ \mathrm{Tr}
\left( m_i \Phi_i^2 \right) + 
\frac{1}{2 k g^2_{\mathrm{YM}}} \int \! d^2 \bar{\theta} \ \mathrm{Tr}
\left( \bar{m}_i \overline{\Phi}_i^2 \right). 
\label{eq:asdef1}
\end{eqnarray}
In terms of 
component fields, this is written as 
\begin{align}
\mathcal{L}_m^{\mathcal{N} = 4} = \mathcal{L}^{\mathcal{N} = 4} +
\frac{1}{k g^2_{\mathrm{YM}}} \mathrm{Tr} \biggl[ &-  | m_i |^2 | 
\phi_i |^2 + \sqrt{2} \varepsilon^{ijk} \bar{m}_i \bar{\phi}_i \phi_j
\phi_k \notag\\
&- \sqrt{2} \varepsilon^{ijk} m_i \phi_i \bar{\phi}_j \bar{\phi}_k
- \frac{1}{2} m_i \psi_i^2 - \frac{1}{2} \bar{m}_i \bar{\psi}_i^2 \biggr].
\end{align}

The deformation 
terms (\ref{N4as1}) are written as
\begin{eqnarray}
\delta \mathcal{L}
&\equiv& - \frac{1}{6} M^{abc} \mathrm{Tr} \left[ 
\frac{}{} \varphi_a \varphi_b \varphi_c \right]
- \frac{1}{2} \mathrm{Tr} \left[ 
m_{AB} \Lambda^{\alpha A} \Lambda_{\alpha} {}^B + m^{AB}
\overline{\Lambda}_{\dot{\alpha} A} \overline{\Lambda}^{\dot{\alpha} } {}_B
\right], 
\end{eqnarray}
where 
\begin{eqnarray}
M^{abc} = m_{AB} (\Sigma^a \overline{\Sigma}^b \Sigma^c )^{AB} 
+ m^{AB} (\overline{\Sigma}^a \Sigma^b \overline{\Sigma}^c)_{AB},
\end{eqnarray}
and $m_{AB}$ and $m^{AB}$ are $4\times 4$ matrices, 
which are  $m_{AB} = 0$, $m^{AB} = - 
\frac{1}{4} C^{(AB)}$ 
in the (A,S)-type deformation
\footnote{
Here we assume the weight factor $\frac{2}{3}$
for the amplitude $\langle \! \langle V_{H_{\varphi \varphi}} V_{\varphi} 
V_{\mathcal{F}} \rangle \! \rangle $, 
which could  be determined by evaluating the five-point amplitude 
$\langle \! \langle V_{\varphi} V_{\varphi} V_{\varphi} V_{\mathcal{F}} 
\rangle \! \rangle$.}

When $m_{AB}$ and $m^{AB}$ 
take a diagonal form
\begin{eqnarray}
& & m_{AB} = \mathrm{diag} ( - m_0, m_1, m_2, m_3), \nonumber \\
& & m^{AB} = \mathrm{diag} ( - \bar{m}_0, \bar{m}_1, \bar{m}_2, \bar{m}_3)
\end{eqnarray}
and $m_0=\bar{m}_0=0$, we find 
\begin{eqnarray}
\delta \mathcal{L} &\equiv& - \frac{1}{6} M^{abc} \mathrm{Tr} \left[ 
\frac{}{} \varphi_a \varphi_b \varphi_c \right]
- \frac{1}{2} \mathrm{Tr} \left[ 
m_{AB} \Lambda^{\alpha A} \Lambda_{\alpha} {}^B + m^{AB}
\overline{\Lambda}_{\dot{\alpha} A} \overline{\Lambda}^{\dot{\alpha} } {}_B
\right] 
\nonumber \\
&=& \mathrm{Tr} \left[ 
\sqrt{2} \varepsilon^{ijk} \bar{m}_i \bar{\phi}_i \phi_j
\phi_k - \sqrt{2} \varepsilon^{ijk} m_i \phi_i \bar{\phi}_j \bar{\phi}_k
- \frac{1}{2} m_i \psi_i^2 - \frac{1}{2} \bar{m}_i \bar{\psi}_i^2
\right]
\end{eqnarray}
which gives the mass deformation ${\cal L}^{{\cal N}=4}_m$.

If we turn on $m_0, \bar{m}_0$, the Lagrangian contains the new terms
\begin{eqnarray}
\delta \mathcal{L}
&=& \mathrm{Tr} \left[ \frac{\sqrt{2}}{3} \varepsilon^{ijk} m_0 (\phi_i \phi_j \phi_k )
- \frac{\sqrt{2}}{3}  \varepsilon^{ijk} \bar{m}_0 (\bar{\phi}_i \bar{\phi}_j \bar{\phi}_k) - \frac{1}{2} m_0 \lambda^2
- \frac{1}{2} \bar{m}_0 \bar{\lambda}^2 \right].
\end{eqnarray}
The first two terms are written as
\begin{eqnarray}
- \frac{1}{k g^2_{\mathrm{YM}}} \frac{\sqrt{2}}{3} \varepsilon^{ijk} \int \! d^2 \theta \ \mathrm{Tr} 
\left[ e^{m_0 \theta^2} \Phi_i \Phi_j \Phi_k \right]
+ \frac{1}{k g^2_{\mathrm{YM}}} \frac{\sqrt{2}}{3} \varepsilon^{ijk} \int \! d^2 \bar{\theta}
\ \mathrm{Tr} \left[ e^{ \bar{m}_0 \bar{\theta}^2} \overline{\Phi}_i \overline{\Phi}_j 
\overline{\Phi}_k \right],
\label{eq:asdef2}
\end{eqnarray}
which are regarded as the deformation of superpotential.
The last two terms are regarded as deformation of gauge coupling constant:
\begin{eqnarray}
\frac{1}{16 k g^2_{\mathrm{YM}}} \mathrm{Tr} \left[
\int \! d^2 \theta \ e^{2 m_0 \theta^2} W^2 + \int \! d^2 \bar{\theta} \
e^{2 \bar{m}_0 \bar{\theta}^2 } \overline{W}^2 \right].
\label{eq:asdef3}
\end{eqnarray}
The Lagrangian (\ref{eq:asdef1}) together with (\ref{eq:asdef2}) and
(\ref{eq:asdef3}) becomes the (A,S)-type deformed one.

\subsection{$\mathcal{N} = 2$ deformation}
We can also write down the deformed ${\cal N}=2$ Lagrangians in ${\cal N}=1$
superspace.
The $\mathcal{N} = 2$ super Yang-Mills theory 
in $\mathcal{N} = 1$ superspace is given by
\begin{eqnarray}
\mathcal{L}^{\mathcal{N} = 2} = \frac{1}{k g^2_{\mathrm{YM}}} \int \! d^2 \theta d^2 \bar{\theta} \
\mathrm{Tr} \left[ \overline{\Phi} e^{2V} \Phi e^{-2V} \right]
+ \frac{1}{16 k g^2_{\mathrm{YM}}} \mathrm{Tr} \left[
\int \! d^2 \theta \ W^{\alpha} W_{\alpha} + \int \! d^2 \bar{\theta}
\ \overline{W}_{\dot{\alpha}} \overline{W}^{\dot{\alpha}} \right],
\nonumber \\
\end{eqnarray}
where $\Phi$ and $\bar{\Phi}$ are chiral and anti-chiral
superfields.

For the (S,A)-type deformation,
the interaction terms (\ref{n2sa}) 
are  written as
\begin{eqnarray}
\mathcal{L}^{(1)}_{\mathrm{(S,A)}} + 
\mathcal{L}^{(2)}_{\mathrm{(S,A)}} 
&=& - \frac{1}{2 k g^2_{\mathrm{YM}}} \int \! d^4 \theta \ 
\theta^2 \bar{\theta}^2 \mathrm{Tr} \left[ \overline{\Phi} D_{\alpha}
W_{\beta} \tilde{C}^{\alpha \beta} \right]
+ \frac{1}{2 k g^2_{\mathrm{YM}}} \int \! d^2 \bar{\theta} \ \bar{\theta}^2
\left[ \overline{\Phi}^2 \tilde{C}^{\mu \nu} \tilde{C}_{\mu \nu}
\right]. \nonumber \\
\end{eqnarray}
Therefore the (S,A)-type deformation in superspace is 
realized by introducing
new interaction term in the D- and F-terms.
It would be interesting to examine this deformation in terms of ${\cal
N}=2$ 
superspace and its relation to the $\Omega$-deformation of ${\cal N}=2$
super Yang-Mills theory \cite{LoMaNe, Shadchin}.

We now discuss the (A,S)-type deformation.
As in the ${\cal N}=4$ case, the (A,S)-type deformation is 
realized by the deformation of coupling parameters.
The mass deformation of ${\cal N}=2$ super Yang-Mills theory is
described by the Lagrangian 
\begin{eqnarray}
\mathcal{L}_m^{\mathcal{N} = 2} &=& \mathcal{L}^{\mathcal{N} = 2}
 + \frac{1}{kg^2_{\mathrm{YM}}} \mathrm{Tr} 
\left[ 
\frac{1}{2} \int \! d^2 \theta \ 
m \Phi^2 + \frac{1}{2} \int \! d^2 \bar{\theta} \
 \bar{m} \overline{\Phi}^2 \right] \nonumber \\
&=& \mathcal{L}^{\mathcal{N} = 2} + \mathrm{Tr} \left[ 
- |m|^2 |\phi|^2 
- \frac{1}{2} m \psi^2 - \frac{1}{2} \bar{m} \bar{\psi}^2 \right].
\end{eqnarray}
Here we have integrated out auxiliary fields 
of the superfields.
If we diagonalize the  background
 $C^{(ij)} = \mathrm{diag} (\bar{m}_0, 
\bar{m})$, the (A,S)-type deformation term (\ref{n2as}) is written as
\begin{eqnarray}
\mathcal{L}_{\mathrm{(A,S)}}^{(1)} = \frac{1}{k g^2_{\mathrm{YM}}}
\mathrm{Tr} \left[ - \bar{m}_0 \bar{\lambda}^2
- \bar{m} \bar{\psi}^2 \right].
\end{eqnarray}
The second term gives mass deformation term with $m=0$.
On the other hand, the first term 
is written in terms of superspace valued gauge coupling as
\begin{eqnarray}
\frac{1}{16 k g^2_{\mathrm{YM}}} \int \! d^2 \bar{\theta} \left[ 
e^{4 \bar{m}_0 \bar{\theta}^2} \overline{W}_{\dot{\alpha}} 
\overline{W}^{\dot{\alpha}}
\right].
\end{eqnarray}

\section{Conclusions and Discussion \label{discussion} }
In this paper we studied the first and second order corrections from
the constant R-R 3-form backgrounds to ${\cal N}=2$ and $4$ super 
Yang-Mills theories, which are realized as 
the low-energy effective field theories on the
(fractional) D3-branes in type IIB superstring theory.
We argued the (S,A) and (A,S)-type R-R backgrounds ${\cal F}$, which correspond
to the R-R (dual) 3-form field strengths in closed superstring
backgrounds.
We also used the scaling condition, where $(\alpha')^{1/2}{\cal F}$ is 
fixed in the zero-slope limit, to calculate the disk amplitudes including
a closed string R-R vertex operator.

The (S,A)-type background with this scaling condition is particularly 
useful to study non-perturbative effects of super Yang-Mills theory.
In fact, the instanton effective action of ${\cal N}=2$ super Yang-Mills
theory with the (S,A)-type deformation agrees with that in
$\Omega$-background at the lowest order in the
deformation parameter and gauge coupling constant \cite{BiFrFuLe}.
The $\Omega$-background is an important setup to applying a localization 
formula to the integration over the instanton moduli space \cite{Ne,
LoMaNe,NeOk, Shadchin}.
It is an interesting problem to extend this correspondence to 
 ${\cal N}=4$ theory or the (A,S)-type deformation 
and examine how the instanton moduli space and the low
energy-effective action are 
deformed by this background 
since the (S,A) and (A,S) deformed $\mathcal{N} = 4$ 
super Yang-Mills
 theories can accommodate both 
self-dual tensor and vector
backgrounds simultaneously from the viewpoint of $\mathcal{N} = 2$ 
deformations. 
This is also important in order to study the nonperturbative 
superstring vacua in the presence of R-R backgrounds.

We examined supersymmetry of the deformed ${\cal N}=4$ action and find that 
${\cal N}=4$ supersymmetry is broken for generic deformation parameter.
But for special case, a part of supersymmetries are unbroken and also
deformed by the R-R background, which are similar to the
non(anti)commutative superspace.
We argued the rank condition for deformation parameter to determine 
unbroken supersymmetries.
Deformations of $\mathcal{N} = 2$ and $\mathcal{N} = 1$
super Yang-Mills theories are described by the orbifold construction,
and the number of unbroken supersymmetries are determined by the
rank condition for deformation parameters.
It would be interesting to study how the central charge of extended
supersymmetry algebra is deformed in these backgrounds.

The (S,A) and (A,S)-type deformations of action cannot be 
realized by the deformation in non(anti)commutative superspace 
\cite{Se} since the spinor index 
structure of the background is different from 
that of  non(anti)commutative superspace. 
In the scaling condition $(\alpha')^{1/2} {\cal F}$ fixed, 
we find that the (A,S)-type deformation of 
${\cal N}=2$ super Yang-Mills theory is realized by allowing 
coupling constants to take values in ${\cal N}=1$ superspace.
For the (S,A) case, we need to introduce further interaction terms
for superfields.
But some interaction terms take simple form when
we use ${\cal N}=2$ extended superspace.
Therefore 
it would be interesting to examine this deformation as the geometry
of ${\cal N}=2$ superspace, as discussed in the case of
$\Omega$-background deformation \cite{NeOk, Shadchin}.

\subsection*{Acknowledgments}
The work of K.~I. is  supported in part by the Grant-in-Aid for Scientific 
Research No. 18540255 from Ministry of Education, Science, 
Culture and Sports of Japan. 
The work of H.~N. is supported by
the Postdoctoral Research Program of Sungkyunkwan University (2007)
and is the result of research activities (Astrophysical Research
Center for the Structure and Evolution of the Cosmos (ARCSEC))
supported by KOSEF.
The work of S.~S. is supported by bilateral exchange program between 
Japan Society for the Promotion of Science (JSPS) and the Academy of Finland.

\end{document}